\documentclass{aa}

\usepackage[varg]{txfonts}
\usepackage{natbib}
\usepackage{hyperref}
\usepackage{graphicx}
\usepackage{caption}
\usepackage{subcaption}
\usepackage{url}
\usepackage{amsmath}
\usepackage{amssymb} 
\usepackage{bm}
\usepackage{amsfonts}
\usepackage{aas_macros}
\hypersetup{colorlinks=true,citecolor=blue,linkcolor=red}
\usepackage[ruled]{algorithm2e}
\usepackage{algorithmic}

\DeclareMathOperator\dom{dom}
\DeclareMathOperator\epi{epi}
\DeclareMathOperator{\prox}{prox}
\DeclareMathOperator{\proj}{proj}
\DeclareMathOperator*{\argmin}{\operatornamewithlimits{arg\,min}}

\title{High Resolution Weak Lensing Mass-Mapping Combining Shear and Flexion} 
\titlerunning{High Resolution Weak Lensing mass-mapping}

\author{F. Lanusse \inst{1,2}\thanks{flanusse@andrew.cmu.edu}, J.-L. Starck \inst{2}, A. Leonard\inst{3}, S. Pires \inst{2}}

\institute{$^1$ McWilliams Center for Cosmology, Department of Physics, Carnegie Mellon University, Pittsburgh, PA 15213, USA.\\
$^2$ Laboratoire AIM, UMR CEA-CNRS-Paris, Irfu, SAp, CEA Saclay, F-91191 GIF-SUR-YVETTE CEDEX, France.\\
$^3$ Department of Physics and Astronomy, University College London, Gower Place, London WC1E 6BT, U.K.}

\begin{document}

\abstract
{} {
We propose a new mass-mapping algorithm, specifically designed to recover small-scale information from a combination of gravitational shear and flexion. Including flexion allows us to supplement the shear on small scales in order to increase the sensitivity to substructures and the overall resolution of the convergence map without relying on strong lensing constraints.
}
{
In order to preserve all available small scale information, we avoid any binning of the irregularly sampled input shear and  flexion fields and treat the mass-mapping problem as a general ill-posed inverse problem, regularised using a robust multi-scale wavelet sparsity prior. The resulting algorithm incorporates redshift, reduced shear, and reduced flexion measurements for individual galaxies and is made highly efficient by the use of fast Fourier estimators.
}
{
We test our reconstruction method on a set of realistic weak lensing simulations corresponding to typical HST/ACS cluster observations and demonstrate our ability to recover substructures with the inclusion of flexion which are lost if only shear information is used. In particular, we can detect substructures  at the 15$^{\prime  \prime}$ scale well outside of the critical region of the  clusters. In addition, flexion also helps to constrain the shape of the central regions of the main dark matter halos. Our mass-mapping software, called Glimpse2D, is made freely available at \url{http://www.cosmostat.org/software/glimpse}.
} {}


\maketitle

\section{Introduction}

Probing the mass distribution of the various structural components of the Universe is a fundamental tool to constrain cosmological models. Gravitational lensing, which is sensitive to the total mass content, is particularly well suited for this task. One of the strengths of gravitational lensing is its ability to probe the dark matter content from the largest cosmological scales down to the galaxy scale, providing valuable constraints at all intermediary scales. 

In particular, at the galaxy cluster scale, gravitational lensing is an established tool for measuring the total mass distribution and investigating the nature of dark matter \citep[e.g.][]{Clowe2006, Massey2015}. While the combination of gravitational shear and strong lensing constraints has proven very successful to resolve the fine structure of galaxy clusters \citep{Bradac2005, Cacciato2006,Merten2009}, strong lensing is only effective within the Einstein radius of the cluster which represents for most clusters only a fraction of their total area. Outside of this region, constraints from gravitational shear are generally very poor below the arcminute scale which makes the detection of substructures extremely unlikely. A promising avenue to bridge the gap between shear and strong lensing, and help resolve substructure outside of the innermost regions of galaxy clusters is gravitational flexion, which has already been shown to provide valuable constraints at these intermediate scales \citep{Leonard2007}. 

Measuring flexion has proven to be a difficult task in practice \citep{Viola2012,Rowe2013}. Originally, flexion estimators were derived from shapelet coefficients \citep{Goldberg2005, Bacon2006}. Another approach relied on moments of galaxy images \citep{Okura2007, Goldberg2007, Schneider2008}. More recently, a different approach to flexion estimation, called Analytic Image Method (AIM), was proposed by \cite{Cain2011} based on forward fitting analytical galaxy image models. Among the benefits of this promising new method are a complete invariance under the mass-sheet degeneracy and a better characterisation of the measurement errors.

Very few methods have been developed for mass-mapping including flexion. The first reconstruction technique using flexion has been proposed in \cite{Bacon2006} as an extension of the Kaiser-Squires Fourier estimator \citep{Kaiser1993}. A parametric reconstruction of Abell 1689 was attempted in \cite{Leonard2007}  and \cite{Leonard2009, Leonard2010} later proposed an extension of aperture mass filters to the case of flexion. Joint reconstruction based on maximum likelihood methods combining shear, flexion and strong lensing have also been proposed by \cite{Er2010} and most recently in \cite{Cain2015}. 

Recovering  high resolution mass-maps from weak lensing is a difficult problem for two main reasons: the irregular sampling of the lensing field and the low Signal to Noise Ratio (SNR) on small scales. While the lensing equations can be most easily inverted when the field is fully sampled on a regular grid, for instance using a Kaiser-Squires estimator, in practice the shear and flexion fields are only sampled at the position of the background galaxies, which are not regularly distributed. In that case the lensing equations are no longer directly invertible and mass-mapping becomes an ill-posed inverse problem i.e. the solution is  not unique and/or extremely  unstable to measurement noise. To avoid this issue, a usual approach is to bin or smooth the input shear or flexion on a regular grid, which not only serves as a regularisation of the inverse  problem but also acts as a denoising step by suppressing smaller scales, where the noise  typically dominates. Although this reasoning holds for shear alone,  adding flexion information greatly increases the SNR of structures on small scales, in which case binning or smoothing the input data can destroy relevant information.

The aim of this paper is to propose a new mass-mapping methodology, specifically designed to recover small-scale information from weak lensing by avoiding any binning or smoothing of the input data and supplementing shear information with flexion to increase the sensitivity to substructures. We address the mass-mapping problem as a general ill-posed  inverse problem which we regularise using a wavelet based sparsity prior on the recovered convergence map. Similar sparse regularisation methods have already been proposed in the context of weak lensing mass-mapping in \cite{Pires2009} for inpainting missing data and \cite{Leonard2012,Leonard2014} for 3D mass-mapping. In this work, the sparse regularisation framework not only allows us to solve the lensing equations even at very low sampling rates, it also provides a very robust multi-scale noise regularisation. Furthermore, despite the irregular sampling of the data, our algorithm relies on fast Fourier estimators to invert the lensing equations, by treating the computation of the Fourier transform (which is ill-posed in this case) as part of the same  inverse problem. This makes our approach far more efficient than the costly finite differencing schemes usually used in maximum likelihood algorithms. Finally, the proposed algorithm accounts for reduced shear and flexion and incorporates individual redshift measurements of background sources. 

This paper is structured as follows. We present a short overview of the weak lensing formalism and motivate the interest of flexion in \autoref{sec:lensing_formalism}.  In \autoref{sec:simplified} we introduce and demonstrate the effectiveness of our sparse regularisation approach in the simplified setting of reconstructing the convergence from shear alone. Then, in \autoref{sec:cluster_mapping}, we incorporate the non-linearity induced by the reduced shear,  we add redshift information for individual galaxies and we extend the framework to include flexion. Finally, in \autoref{sec:sim_results} we demonstrate on realistic cluster simulations that our method is successful at reconstructing the surface mass density and  illustrate the impact of the additional flexion information for the recovery of small scale substructures. 

\section{Weak gravitational lensing formalism} 
\label{sec:lensing_formalism}
\subsection{Weak lensing in the linear regime}

Because gravitational lensing is sensitive to the total matter content of structures, it is an ideal probe to map the dark matter distribution. The presence of a massive lens along the line of sight will induce a distortion of the background galaxy images which can be described by a coordinate transformation \citep{Bartelmann2010} between the unlensed coordinates $\bm{\beta}$ and the observed image coordinates $\bm{\theta}$:
\begin{equation}
	\bm{\beta} = \bm{\theta} - \nabla \psi(\bm{\theta}) \;,
	\label{eq:coordinates}
\end{equation}
where $\psi$ is the 2D lensing potential encoding the deflection of light rays by the gravitational lens. The lensing potential can be seen as generated from a source term, through the following Poisson equation:
\begin{equation}
\Delta \psi(\bm{\theta}) = 2 \ \kappa(\bm{\theta}) \;,
\end{equation}
where $\kappa$ is the convergence, a dimensionless surface density. If we consider a single thin lens, $\kappa$ can be derived from the surface mass density of the lens $\Sigma$ as:
\begin{equation}
	\kappa(\bm{\theta}) = \frac{\Sigma(\bm{\theta})}{\Sigma_{\mathrm{critic}}} \;,
\end{equation}
where $\Sigma_{\mathrm{critic}}$ is the critical surface density defined by a ratio of distances between the observer, the lens and the source:
\begin{equation}
\Sigma_{\mathrm{critic}} = \frac{c^2}{4 \pi G} \frac{D_S}{D_L D_{LS}}\;.
\end{equation}
In this expression, $D_S$, $D_L$ and $D_{LS}$ are respectively the angular diameter distances to the source, the lens and between the lens and the source.

If the lensing effect is weak enough, the coordinate transformation can be approximated using a Taylor expansion to first order of \autoref{eq:coordinates} as:
\begin{equation}
	\beta_i \simeq A_{i j} \theta_j\;,
\end{equation}
where $\mathbf{A}$ is the amplification matrix, defined in terms of the derivatives of the lensing potential and which takes the form:
\begin{align}
	A_{i j}(\bm{\theta}) &= \frac{\partial \beta_i}{\partial \theta_j} = \delta_{i j} - \partial_i \partial_j \psi(\bm{\theta}) \;,\\
	\mathbf{A}	&= \left(\begin{matrix}
	1 - \kappa - \gamma_1 &  - \gamma_2 \\
	  - \gamma_2          & 1 - \kappa + \gamma_1
	\end{matrix}\right) \;, 
\end{align}
where $\gamma$ is the complex shear, causing anisotropic distortions whereas $\kappa$ only produces an isotropic magnification effect. 
However, in practice, when using galaxy ellipticities as a measurement of the lensing effect, neither the convergence nor the shear are directly observable. Indeed, the net contribution of weak lensing to the ellipticity of a galaxy is not the shear $\gamma$ but the reduced shear $g$ defined as $g = \frac{\gamma}{1 - \kappa}$. The amplification matrix can equivalently be expressed as a function of the reduced shear by factorising a term $(1 - \kappa)$ which yields:
\begin{equation}
	\mathbf{A}	= (1 - \kappa) \left(\begin{matrix}
	1 - g_1 &  - g_2 \\
	  - g_2    & 1 + g_1
	\end{matrix}\right) \;, 
\end{equation}
Note that in the weak lensing regime, for $\kappa \ll 1$, the reduced shear can be approximated to the actual shear, $\gamma \simeq g$. However, this assumption no longer holds for the purpose of mapping galaxy clusters where it is critical to correctly account for the reduced shear. 


\subsection{Gravitational flexion}
\label{sec:flexion}

The previous expressions for the coordinate transformations hold if the convergence and shear fields are assumed to be constant at the scale of the observed source images. When this assumption is not verified, the expansion of the coordinate transformation can be pushed to second order \citep{Goldberg2005}:
\begin{equation}
	\beta_i \simeq A_{ij} \theta_j + \frac{1}{2} D_{i j k} \theta_j \theta_k \;,
\end{equation}
where the third order lensing tensor $D_{i j k}$ can be derived from $D_{i j k} = \partial_k A_{ij}$. This additional term gives rise to third order derivatives of the lensing potential which can be summarised as a spin-1 field $\mathcal{F} = \nabla \kappa $ and a spin-3 field $\mathcal{G} = \nabla \gamma$ respectively called first and second flexion.

Similarly to the case of the reduced shear, these two flexion components  are not directly observable. Instead, only the reduced flexion fields are measured from galaxy images:
\begin{equation}
	F = \frac{\mathcal{F}}{ 1- \kappa} \quad \mbox{ ; } \quad G = \frac{\mathcal{G}}{ 1- \kappa} \;.
\end{equation}
Furthermore, estimating in practice the second flexion $G$ is extremely delicate and measurements are generally noise dominated and discarded. As a result, in the rest of this work, we will only include measurements of the reduced first flexion $F$. 

Throughout this work, we assume reduced shear and flexion measurement provided by the AIM method of \cite{Cain2011}. In particular, we will use $\sigma_F = 0.029$ arcsec$^{-1}$ for the dispersion of flexion measurements, as reported by \cite{Cain2011} on HST data.

\subsection{Mass-mapping using Fourier estimators}
\label{sec:Fourier_estimator}

The most well known inversion technique for weak lensing mass-mapping is the Kaiser-Squires inversion \citep{Kaiser1993} which consists in applying a minimum variance filter in Fourier space to both components of the shear, yielding one estimate of the convergence field with minimum noise variance, given by
\begin{equation}
\hat{ \tilde{\kappa}}_\gamma = \frac{k_1^2 - k_2^2}{k^2}\tilde{\gamma}_1 + \frac{2 k_1 k_2}{k^2}\tilde{\gamma}_2\;,
\label{eq:shear_minimum_variance}
\end{equation}
where $k^2 = k_1^2 +k_2^2$. This estimator is only valid for $k^2 \neq 0$ and as a result the mean value of the field cannot be recovered, which corresponds to the well-known mass sheet degeneracy. Assuming white Gaussian noise of variance $\sigma_{\gamma}^2$ on the input shear field, the Kaiser-Squires estimator features a simple flat noise power spectrum:
\begin{equation}
		< \hat{\tilde{\kappa}}_{\gamma}^* \hat{\tilde{\kappa}}_{\gamma} > = \sigma_{\gamma}^2
\end{equation}
Although this inversion technique has some very desirable properties including linearity and minimum variance, it has one major shortcoming, namely that is not well defined on bounded domains, even less so on domains with complex geometries and missing data, which occurs in actual surveys. Some alternatives to the simple Kaiser-Squires have also been developed to mitigate some of it shortcomings \citep{Pires2009,Deriaz2012} using iterative schemes. 

Similarly to the Kaiser-Squires inversion for the shear, \cite{Bacon2006} proposed a minimum variance filter for estimating the convergence from first flexion $\mathcal{F}$. As the flexion is only the gradient of the convergence, these two quantities can be related in the Fourier domain as:
\begin{equation}
	\tilde{\mathcal{F}}_1 = - i k_1 \tilde{\kappa} \quad ; \quad \tilde{\mathcal{F}}_2 = -i k_2 \tilde{\kappa} \;.
\end{equation}
With the same approach as the Kaiser-Squires inversion, the two components of the first flexion can be combined in order to yield a minimum variance estimator of the convergence, which takes the simple expression:
\begin{equation}
\hat{\tilde{\kappa}}_F = \frac{i k_1 \tilde{\mathcal{F}}_1 + ik_2 \tilde{\mathcal{F}}_2}{k^2}\;.
\end{equation}
Note that, just as with the Kaiser-Squires inversion, this expression is only valid for $k_1 \neq 0$ and $k_2 \neq 0$, which means that the flexion reconstruction is still subject to the mass sheet degeneracy. Contrary to the previous Kaiser-Squires inversion, this estimator does not have a flat noise power spectrum. Indeed, assuming uncorrelated flexion measurements with intrinsic variance $\sigma_F^2$, the variance of this estimator is:
\begin{align}
	< \hat{\tilde{\kappa}}_F^* \hat{\tilde{\kappa}}_F > &=  \frac{k_1^2}{k^4} \sigma_F^2 + \frac{k_2^2}{k^4} \sigma_F^2  \;,\\
	&= \frac{1}{k^2} \sigma_F^2 \;.
\end{align}
This makes the reconstruction of mass maps from flexion alone using this estimator very problematic on large scales where the noise will always dominate the signal. This means that applying any low-pass smoothing actually reduces the signal to noise ratio (SNR) of the reconstruction.

\begin{figure}[h]
\centering
\includegraphics[width=\columnwidth]{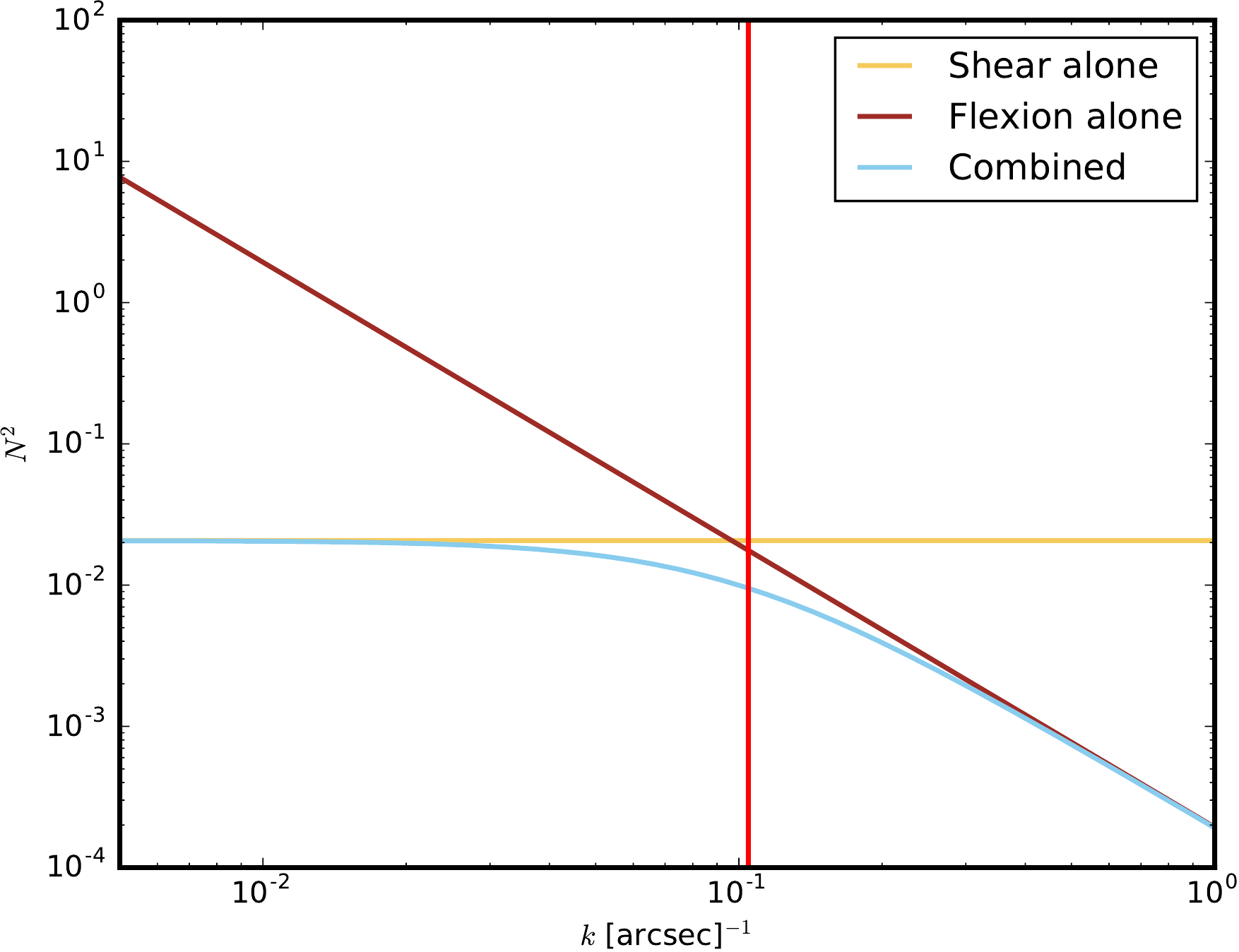}
\caption[Noise power spectra of the minimum variance estimators for shear, flexion and shear and flexion combined.]{Noise power spectra of the minimum variance estimators for shear alone, flexion alone and shear and flexion combined. The vertical red line indicates the arcminute scale. These noise power spectra assume $\sigma_F=0.029$ arcsec$^{-1}$, $\sigma_\gamma=0.3$ and $n_g = 80$ gal/arcmin$^2$ with 14 arcsecs pixels.}
\label{fig:flexionVSshearPS}
\end{figure}

 \autoref{fig:flexionVSshearPS} shows the noise power spectra of a shear inversion alone and flexion inversion alone, the arcminute scale is marked by the vertical line. As can be seen, the flexion noise power-spectrum is much more favourable  than the shear on small scales. However, it is only worth considering adding flexion information if the scale at which these two power spectra cross is still relevant. This scale only depends on the ratio of shear and flexion intrinsic dispersion and we find it to be around 65 arcsecs for  $\sigma_F=0.029$ arcsec$^{-1}$ and $\sigma_\gamma=0.3$. As a result, based solely on considerations about the relative noise power spectra using these Fourier estimators, we argue that flexion can be used to help resolve sub-arcmins structures.

It is clear from \autoref{fig:flexionVSshearPS} that although flexion can help constrain small features, it is not competitive with respect to the shear on scales larger than 1 arcmin. To simultaneously benefit from both shear and flexion, \cite{Bacon2006} proposed a minimum variance filter combining both measurements. Although they present reconstructions using this combined filter, they do not explicitly provide its expression, which can easily be derived and takes the form:
\begin{equation}
	\hat{\tilde{\kappa}}_{\gamma F} =  \frac{1}{k^2 + \frac{\sigma_F^2}{\sigma_\gamma^2}} \left( i k_1 \tilde{\mathcal{F}}_1 + i k_2 \tilde{\mathcal{F}}_2 + \frac{\sigma_F^2}{\sigma_\gamma^2}\left( \frac{k_1^2 - k_2^2}{k^2} \tilde{\gamma}_1 + \frac{2k_1 k_2}{k^2}   \tilde{\gamma}_2\right)\right) \;.
\label{eq:flexion_minimum_variance}
\end{equation}
The noise variance of this estimator is now:
\begin{equation}
	< \hat{\tilde{\kappa}}_{\gamma F}^* \hat{\tilde{\kappa}}_{\gamma F} > = \frac{\sigma_F^2}{k^2 + \frac{\sigma_F^2}{\sigma_\gamma^2}} \;.
\end{equation}
As can be seen, on large scales, i.e. for small $k$ we recover asymptotically the flat shear noise power spectrum while on small scales we recover the  flexion estimator noise power spectrum. \autoref{fig:flexionVSshearPS} illustrates the noise power spectrum of this combined estimator (blue line) in realistic conditions. We see that the combined estimator starts improving over the shear alone around the arcminute scale.

\begin{figure}
\centering
\begin{subfigure}[t]{0.49\columnwidth}
\includegraphics[width=\textwidth]{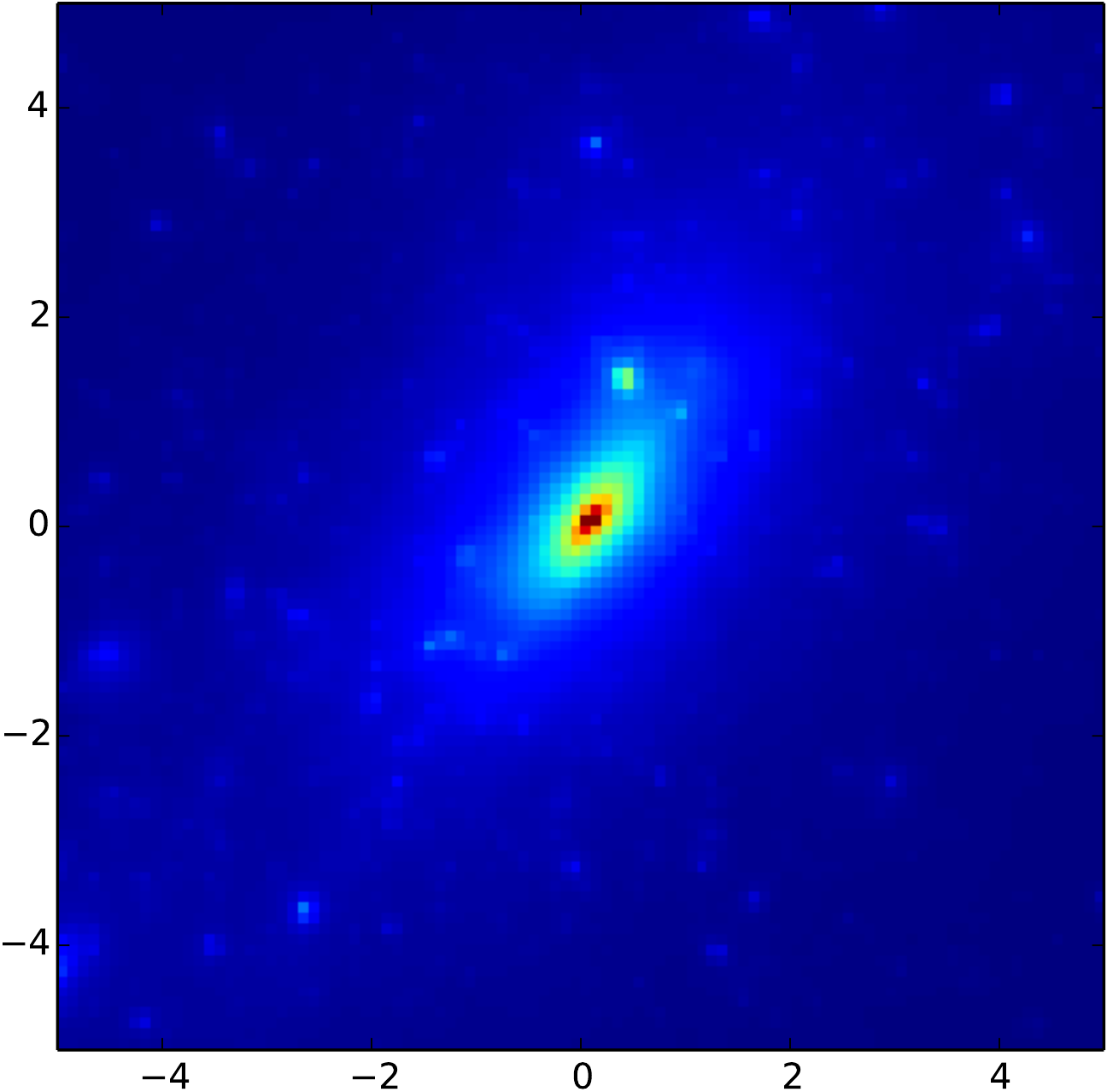}
\caption{Input convergence map}
\end{subfigure}~%
\begin{subfigure}[t]{0.49\columnwidth}
\includegraphics[width=\textwidth]{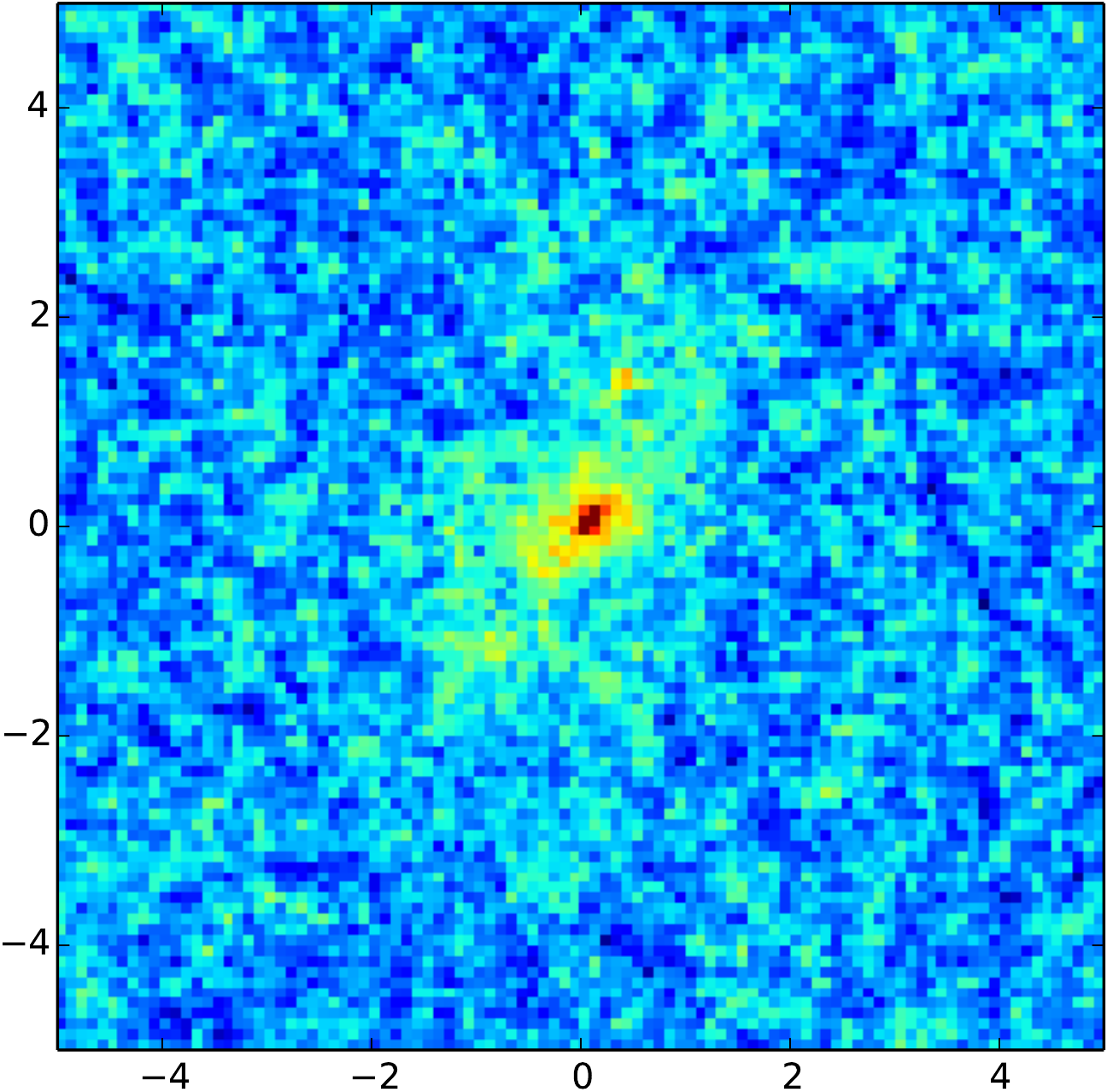}
\caption{Combined estimator $\hat{\kappa}_{\gamma F}$}
\end{subfigure}\\
\begin{subfigure}[t]{0.49\columnwidth}
\includegraphics[width=\textwidth]{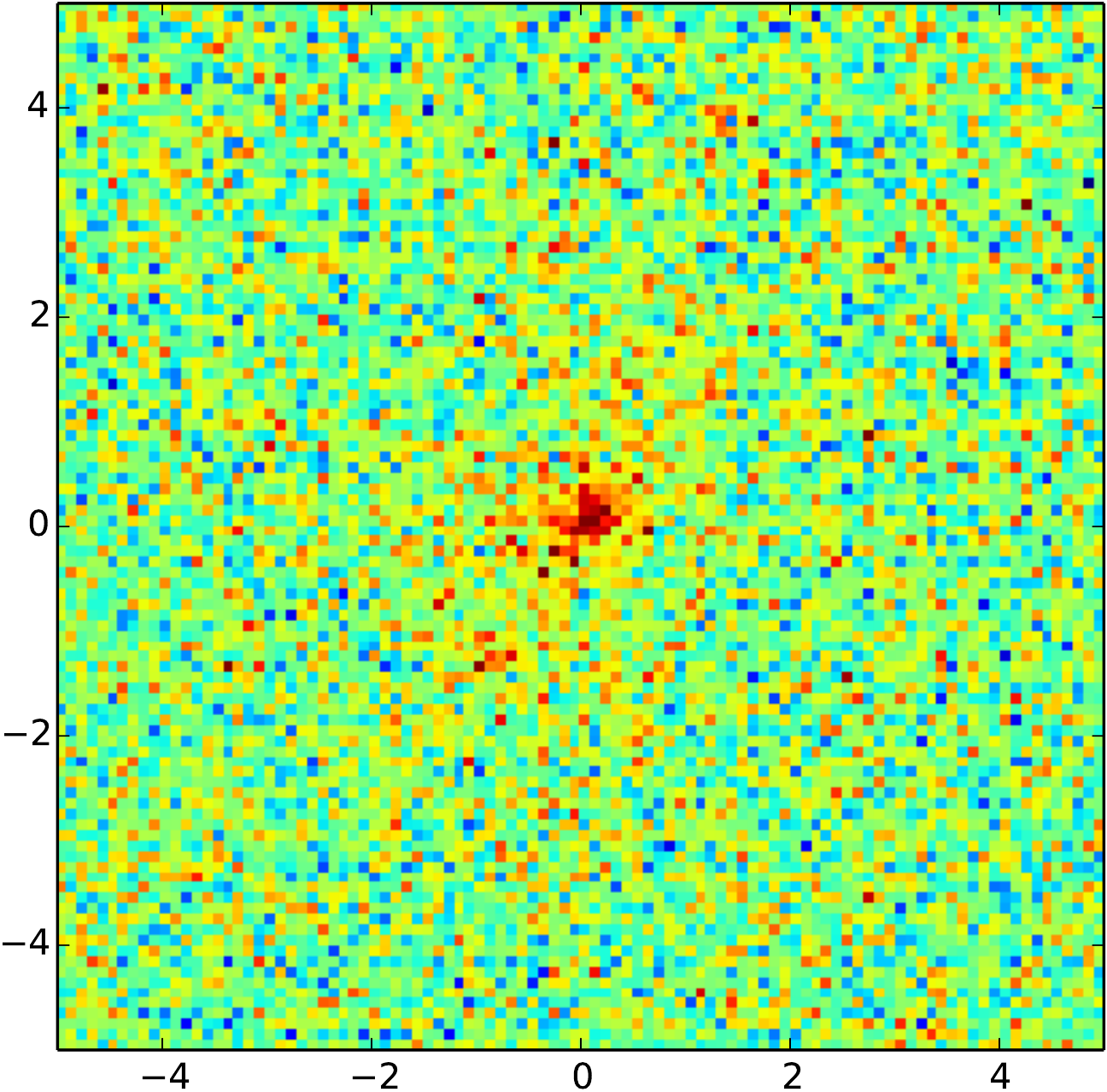}
\caption{Shear alone $\hat{\kappa}_{\gamma}$}
\end{subfigure}~%
\begin{subfigure}[t]{0.49\columnwidth}
\includegraphics[width=\textwidth]{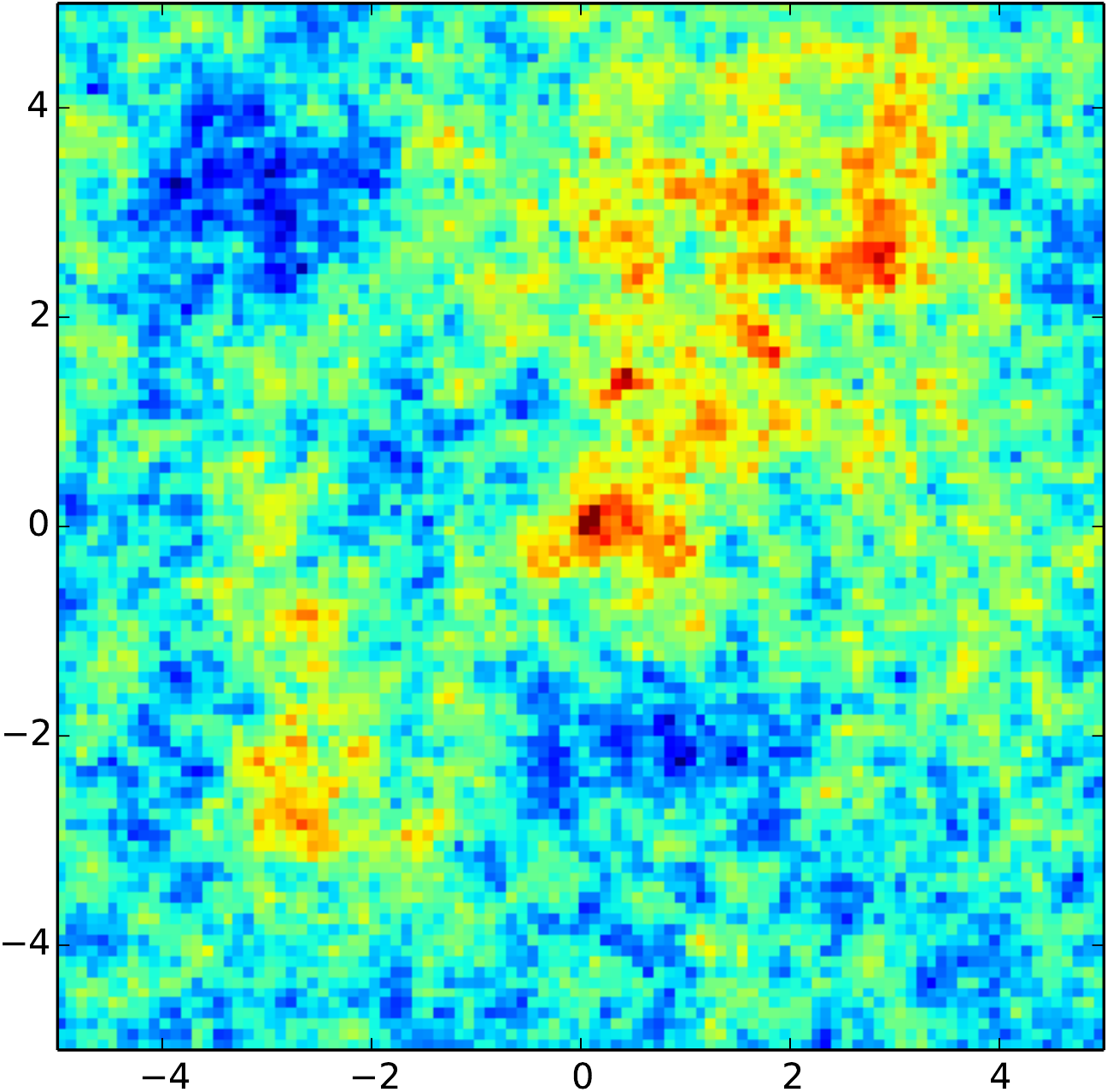}
\caption{Flexion alone $\hat{\kappa}_{F}$}
\end{subfigure}
\caption[Reconstruction of a small $10 \times 10$ arcmin$^2$ field from shear and flexion.]{Top: Simulated $10 \times 10$ arcmin$^2$ field containing a massive cluster (a) and its reconstruction using the combined minimum variance estimator (b). Bottom: Reconstructions using the shear alone Kaiser-Squires estimator (c) and the flexion alone minimum variance estimator (d).The noise level correspond to $\sigma_F=0.04$ arcsec$^{-1}$, $\sigma_\gamma=0.3$ and $n_g = 50$ gal/arcmin$^2$ with 6 arcsecs pixels.}
\label{fig:combined_rec}
\end{figure}

An example of reconstruction using this combined estimator is shown on \autoref{fig:combined_rec}. The input convergence map is reconstructed from shear alone, flexion alone and combining shear and flexion. As expected, the shear alone reconstruction is very noisy, especially on small scales because of its flat noise power spectrum. On the contrary, the flexion alone reconstruction is noise dominated on scales larger than 1 arcmin. However, by combining both shear and flexion information, the noise is effectively suppressed on small scales and not amplified on large scales which makes the cluster and some of its substructure clearly identifiable without any additional filtering. Note however that this simple example only illustrates the noise properties of these estimators but is unrealistic in the sense that it does not consider the problem of missing data.

These Fourier estimators are only valid in the weak lensing regime and should only be applied when the convergence is weak (when $\kappa \ll 1$) such that the observed reduced shear and flexion can be approximated to the true shear and flexion:
\begin{align}
g = \frac{\gamma}{ 1 - \kappa} \simeq \gamma  \quad \mbox{ and } \quad F = \frac{\mathcal{F}}{ 1 - \kappa} \simeq \mathcal{F}
\end{align}
Of course this assumption no longer holds at the close vicinity of galaxy clusters and it is therefore important to properly account for reduced shear and flexion for the purpose of mapping galaxy clusters. In those situations, $\kappa$ is no longer negligible and the reconstruction problem becomes non-linear. As was noted in \cite{Seitz1995}, it is still possible to recover the convergence using the previous linear estimators by iteratively solving the inversion problem, using at each iteration the previous estimate of the convergence to correct the measured reduced shear:
\begin{equation}
	\kappa_{n+1}(\bm{\theta}) -\kappa_0 = \frac{1}{\pi} \int \mathrm{d} \bm{\theta}^\prime (1 - \kappa_n(\bm{\theta}^\prime))  \Re \left[ D^*(\bm{\theta} - \bm{\theta}^{\prime})  g(\bm{\theta}^\prime) \right]
	\label{eq:reducedShearCorrection}
\end{equation}
Where $D(\bm{\theta})$ is the Kaiser-Squires convolution kernel and the constant $\kappa_0$ accounts for the fact that the constant of the convergence is not constrained by the data. This iterative process is generally found to converge quickly to the solution. Of course, a similar procedure can be implemented to correct the reduced flexion and we will use the same principle in the reconstruction algorithm presented in the next section.

\subsection{Mass-sheet degeneracy and redshift dependency}
\label{sec:mass_sheet}

One of the most notorious issues in weak lensing mass-mapping is the so-called mass-sheet degeneracy \citep{Seitz1996}. This degeneracy is due to the fact that the shear is left invariant by the addition of a constant mass-sheet to the lens surface density. As a result, the observed reduced shear is invariant under the following \textit{$\lambda$-transformation}:
\begin{equation}
\kappa^\prime = \lambda \kappa + ( 1 - \lambda)
\end{equation}
Indeed, if $g^\prime$ is the reduced shear generated by $\kappa^\prime$ then:
\begin{equation}
g^\prime = \frac{\lambda \gamma}{1 - \lambda \kappa - 1 + \lambda  } = \frac{\lambda \gamma}{\lambda- \lambda \kappa } = g
\end{equation}
By the same mechanism, the reduced flexion $F$ is also invariant under the same $\lambda$-transformation: adding the flexion information will not affect the mass-sheet degeneracy.

This issue is particularly problematic for measuring the mass of galaxy clusters from weak-lensing. Indeed, these measurements are typically performed on small fields where it cannot simply be assumed that the convergence goes to zero outside the field without biasing the measurement \citep{Bartelmann1995}.

However, although it is true that from shear alone the mass-sheet degeneracy can not be lifted, it can nonetheless be mitigated when including additional information about the relative distances between the lens and the different sources. Indeed, as was pointed out in \cite{Bradac2004}, knowledge of individual photometric redshifts of background galaxies is enough to constrain the mass-sheet for strong enough lenses.

Let us consider $\gamma_\infty(\theta)$ and $\kappa_\infty(\theta)$ the shear and convergence of a given lens at redshift $z_l$, for sources at infinite redshift. The actual shear and convergence applied to a galaxy at a specific redshift $z_s$ can be expressed as $\kappa(\theta, z_s) = Z(z_s) \kappa_\infty(\theta)$ and $\gamma(\theta,z_s) = Z(z_s) \gamma_\infty(\theta)$ where $Z$ is a cosmological weight, function of the redshifts of the source and lens, which can be expressed as a ratio of two critical surface densities:
\begin{equation}
	Z(z_s) = \frac{\Sigma_{crit}^{\infty}}{\Sigma_{crit}(z_s)} H(z_s - z_l)
\end{equation}
where $H$ is the Heaviside step function, which only accounts for the fact that sources located at lower redshift than the lens are not lensed, and $\Sigma_{crit}^\infty = \lim\limits_{z \rightarrow \infty} \Sigma_{crit}(z)$ is the critical surface density for sources at infinite redshift.

 If we now consider the reduced shear measured on a galaxy at a specific redshift $z_s$ as a function of $\kappa_\infty$ and $\gamma_\infty$, it takes the form:
\begin{equation}
	g(\theta, z_s) = \frac{Z(z_s)  \gamma_\infty(\theta)}{1 - Z(z_s) \kappa_\infty(\theta)}
\end{equation}
This expression alone is just a rewriting of the measured reduced shear, which makes explicit the dependency on the redshift of the source. Therefore, the convergence is still subject to the mass-sheet degeneracy which can now be made explicit in term of $z_s$:
\begin{equation}
\kappa^\prime = \lambda \kappa + \frac{1- \lambda}{Z(z_s)}
\end{equation}
It is easily shown that such a transformation leaves the measured reduced shear invariant. However, it is specific to a given source redshift $z_s$ and the reduced shear measured on a galaxy at a different redshift $z_s^\prime \neq z_s$ is not invariant under the same transformation. As a result, simultaneously constraining the convergence to fit the reduced shear at different redshifts formally leads to $\lambda=1$ and breaks the degeneracy.
In practice however, this mechanism is only effective at lifting the degeneracy in regions where the convergence is strong enough. Indeed, in the linear regime the transformation $\kappa^\prime = \kappa + \lambda$ has no dependency on the redshift of the source. Therefore, the mass-sheet degeneracy can only be lifted in the neighborhood of strong enough lenses, where the weak lensing approximation no longer holds. 

For the rest of this work, we will assume knowledge of the redshift $z_L$ of the lens as well as individual photometric redshifts for the sources. For a source with a given photo-z distribution $p_i(z)$, we define the following lensing weight $Z_i$ relating the convergence of the source and the convergence at infinite redshift:
\begin{equation}
	Z_i = \int_{z_L}^\infty Z(z) p_i(z) dz  = \int_{z_L}^\infty \frac{\Sigma_{critic}^{\infty}}{\Sigma_{critic}(z)} p_i(z) dz
\end{equation}
where $\Sigma_{critic}(z)$ is the critical density for a lens at redshift $z_L$ and a source at redshift $z_s = z$ while $\Sigma_{critic}^\infty$ is the critical density for sources at infinite redshift. With this definition, the reduced shear $g_i$ and reduced flexion $F_i$ for a given source can be described as a function of the shear, flexion and convergence for sources at infinite redshift $\gamma_\infty$, $F_\infty$ and $\kappa_\infty$:
\begin{equation}
g_i = \frac{Z_i \gamma_\infty}{1 - Z_i \kappa_\infty}  \quad ; \quad F_i = \frac{Z_i F_\infty}{1 - Z_i \kappa_\infty}  
\end{equation}
Finally, for convenience in the rest of the paper, we introduce the diagonal matrix $\mathbf{Z} = \mathrm{diag}(Z_1, Z_2, \ldots, Z_N )$, with $N$ the number of galaxies in the survey.

\section{A new mass-mapping methodology}
\label{sec:simplified}

\subsection{Dealing with irregularly sampled data}

The global inversion methods presented in the previous section all assume knowledge of the shear on a grid. However, the shear is only sampled in practice at randomly distributed galaxy positions. Therefore a first step in all of these methods is to bin the shear catalogue in order to have a sufficient number of galaxies per pixel prior to performing the inversion. Even so, some regions are bound to remain empty because of various masks applied to the data.
The simplest and most common procedure to handle missing data consists in applying a large Gaussian smoothing to the binned shear maps with a kernel larger than the typical size of the masks. Obviously, this strategy is far from optimal as it irremediably destroys small scale information.

A more sophisticated approach was proposed in \cite{Pires2009}, based on sparse inpainting. This method regularises the inversion in the case of missing data by imposing on the solution a sparsity prior in a Discrete Cosine Transform (DCT) dictionary. In effect, the resulting inpainted map features minimal B-modes leaking and unbiased power spectrum and bispectrum with respect to the original map. This method as been employed to map galaxy clusters in \cite{Jullo2014} where the authors advocate using a very fine binning  of the data in order to have on average one galaxy per pixel.

We propose in this section a new method for reconstructing the convergence map, based on Fourier estimators, which does not require any binning or smoothing of the input shear. Although this approach allows us to preserve small-scale information which would otherwise be lost, it does turn the reconstruction problem into an ill-posed inverse problem, which we will address using sparse regularisation

\subsubsection{Non-equispaced Discrete Fourier Transform}

When using Fourier based methods, the fundamental motivation behind binning or smoothing is to regularise the Discrete Fourier Transform (DFT) of the shear, which is not well defined for data sampled on an irregular grid and which is known in this case as a Non-equispaced Discrete Fourier Transform (NDFT). The NDFT is of course no longer an orthogonal transform and is usually not even invertible. We will consider the case where only the spatial nodes $\bm{x} = \left( x_l \right)_{0 \leq l < M}$ are arbitrary while the frequency nodes $\bm{k}= \llbracket 0, N \llbracket$ are $N$ regularly spaced integers. Computing the NDFT from a set of Fourier coefficients $\hat{\bm{f}}= \left(\hat{f}_{k} \right)_{0 \leq k < N}$ simply amounts to evaluating the trigonometric polynomials:
\begin{equation}
	\forall l \in \llbracket 0, M \llbracket, \quad f_{l} =  \frac{1}{\sqrt{N}} \sum_{k=0}^{N-1} \hat{f}_{k} e^{2 \pi i k x_l} \;.
	\label{eq:NDFT}
\end{equation}
This operation can more conveniently be expressed using matrix notations as:
\begin{equation}
	\bm{f} = \mathbf{T} \hat{\bm{f}} \quad \mbox{ with } \quad T_{l k} =  \frac{1}{\sqrt{N}} e^{2 \pi i k x_l} \;,
\end{equation}
where $\mathbf{T}$ is the NDFT matrix. Note that in the case of equispaced spatial nodes such that $x_l = \frac{1}{N} l$, this operation corresponds to the conventional DFT and $\mathbf{T}$ reduces to the Fourier matrix $\mathbf{F}$ defined as:
\begin{equation}
	F_{l k} = \frac{1}{\sqrt{N}} e^{2 \pi i k l / N} \;.
\end{equation}
This Fourier matrix is unitary and its inverse is simply its Hermitian conjugate: $\mathbf{F}^{-1} = \mathbf{F}^*$. 
On the contrary, for non-equispaced spatial nodes $\bm{x}$, the operator $\mathbf{T}$ is typically neither orthogonal nor admits an inverse. Still, one can consider the adjoint NDFT operator $\mathbf{T}^*$: 
\begin{equation}
	 T^*_{k l} = \frac{1}{\sqrt{N}} e^{-2 \pi i k x_l / N} \;.
\end{equation}
Although this adjoint operation no longer corresponds to the inverse of the transform, it can be used in practice to estimate the inverse when the problem remains over-determined through a least squares optimisation of the form:
\begin{equation}
	\hat{\bm{f}} = \argmin_{\hat{\bm{x}}} \parallel \bm{f} - \mathbf{T} \hat{\bm{x}} \parallel_2^2 \;.
	\label{eq:LS}
\end{equation}
This problem can be efficiently solved using iterative algorithms (in particular using a conjugate gradient) which involve the computation of both $\mathbf{T}$ and $\mathbf{T}^*$. 

It is therefore important to have fast algorithms for the computation of the NDFT and its adjoint. Note that a naive evaluation of the sum in \autoref{eq:NDFT} would scale as $\mathcal{O}(N \times M)$ which is prohibitively large for most applications (including for our mass-mapping problem). In this work, we use a fast approximate algorithm\footnote{The C++ library NFFT 3 is available at \url{https://www-user.tu-chemnitz.de/~potts/nfft}} to evaluate the NDFT and its adjoint, called NFFT \citep{Keiner2009}, which only scales as $\mathcal{O}(N \log(N) + |\log(\epsilon)|M)$ where $\epsilon$ is the desired accuracy. For a given sampling of the frequency space (i.e. for a given $N$), the NFFT only linearly scales with the number $M$ of spatial nodes $x_l$, which in our case will correspond to the number of galaxies in the survey.

Although the  DFT can still be estimated for irregularly sampled data by solving \autoref{eq:LS} when the problem remains over-determined, the  under-determined problem is ill-posed and cannot be directly solved using a least-squares but requires an additional regularisation. 

\subsubsection{Sparse regularisation of inverse problems}
\label{sec:sparse_reg}

For the specific application of weak lensing mass-mapping, the inverse problem of estimating the Fourier transform of the signal for irregular samples can be severely ill-posed when the galaxy density is low. In that case, a simple least-squares such as \autoref{eq:LS} is not able to recover a satisfying estimate of $\hat{f}$ without additional prior. To regularise this inversion, we propose in this paper to use a so-called analysis-based sparsity prior. We introduce here  the concepts of sparse regularisation in a general setting before specialising them to the weak lensing mass-mapping problem in the next section.

Let us consider a general linear inverse problem of the form:
\begin{equation}
 \bm{y} = \mathbf{A} \bm{x} + \bm{n}
 \label{eq:inv_prob}
\end{equation}
where $\mathbf{A}$ is a linear operator, $\bm{y}$ are the measurements, contaminated by an additive Gaussian noise $\bm{n}$, and $\bm{x}$ is the unknown signal we wish to recover from the measurements $\bm{y}$. Ill-posed inverse problems typically occur when the operator $\mathbf{A}$ is not invertible or extremely badly conditioned, in which case the solution may not be unique and is at least extremely unstable to measurement noise. In order to recover a meaningful estimate of the unknown signal some additional prior information is required to help restrict the space of possible solutions. One particularly powerful instance of such a prior is the sparsity prior which assumes that the signal can be represented in an suitable domain using only a small number of nonzero coefficients \citep{Starck2015}. 

More formally, let us consider a signal $\bm{x}$, it can be analysed using a dictionary $\bm{\Phi}$ to yield a set of analysis coefficients $\bm{\alpha}$, such that $\bm{\alpha} = \Phi^* \bm{x}$. For instance, in the case of the Fourier dictionary, $\bm{\Phi}$ would correspond to the DFT matrix  $\bm{F}$ and $\alpha$ would be the  Fourier coefficients of the signal $\bm{x}$. A signal is considered sparse in this analysis framework if only a small number of its coefficients $\bm{\alpha}$ are non-zero.

Under a sparsity prior, the solution of the inverse problem stated in \autoref{eq:inv_prob} can be recovered as the sparsest possible signal $\bar{\bm{x}}$ still compatible with the observations.  This can be formalised as a convex optimisation problem of the form:
\begin{equation}
	\bar{\bm{x}} =  \argmin\limits_{\bm{x}} \frac{1}{2} \parallel \bm{y} - \mathbf{A} \bm{x} \parallel_2^2 + \lambda \parallel \mathbf{\Phi}^* \bm{x} \parallel_1 \;.
	\label{eq:analysis_prior}
\end{equation}
The first term in this expression in this minimisation problem is a quadratic data  fidelity term while the second term promotes the  sparsity of the solution by penalising the $\ell_1$ norm of the analysis  coefficients  of $\bm{x}$.  

Thanks to very recent advances in the field of convex optimisation and proximal theory, we now have at our disposal very efficient algorithms for solving this optimisation  problem, which had remained for a long time largely intractable. In the application presented in this paper, we chose to implement an adaptation of the primal-dual algorithms introduced in \cite{Condat2013, Vu2013}. We direct the interested reader to \autoref{sec:appendix_prox} for a few elements of proximal calculus and details concerning this specific algorithm. A more in depth introduction to proximal theory can be found in \cite{Starck2015}.

\subsection{Sparse recovery algorithm for weak lensing mass-mapping}

We now consider the specific case of weak lensing mass-mapping from irregularly sampled data and we propose a sparse recovery algorithm to reconstruct the convergence map from noisy shear data. 

Consider a lensing survey with $N_g$ galaxies. We can write the expression of the shear $\bm{\gamma}= \left( \gamma_i \right)_{i \in \llbracket 0, N_g \llbracket}$ at the position of each galaxy given a convergence map $\bm{\kappa}$ as
\begin{equation}
	\bm{\gamma} = \mathbf{T} \mathbf{P} \mathbf{F}^* \bm{\kappa}
	\label{eq:forward_mass}
\end{equation}
In this expression, $\mathbf{F}$ is the Fourier matrix and $\mathbf{T}$ is  the NDFT matrix defined for arbitrary spatial nodes $\bm{x}$ placed at the position of each galaxy in the survey. The diagonal operator $\mathbf{P}$ implements the transformation from convergence to shear in Fourier space:
\begin{equation}
	\hat{\bm{\gamma}} = \mathbf{P} \hat{\bm{\kappa}} = \left( \frac{k_1^2 - k_2^2}{k^2}  + i \frac{2 k_1 k_2}{k^2} \right) \hat{\bm{\kappa}}
	\label{eq:forward_mass_fourier}
\end{equation}
Of course, this expression is not defined for $k_1 = k_2 = 0$, which corresponds to the well known mass-sheet degeneracy and by convention we will set the mean to 0. We are using complex notations for both shear and convergence, with in particular $\bm{\kappa} = \bm{\kappa_E} + i \bm{\kappa_B}$ where $\kappa_E$ and $\kappa_B$ are respectively E- and B-modes maps. 

This expression is only formally correct if the Fourier transform is evaluated on the $\mathbb{R}^2$ plane. In practice however, we consider finite fields and rely on a DFT to compute the Fourier transform which imposes artificial periodic conditions. The result of \autoref{eq:forward_mass_fourier} is therefore a circular convolution of the convergence field and the Kaiser-Squires kernel. In practice, when evaluating this expression, we apply sufficient zero-padding to the convergence field $\bm{\kappa}$ in order to minimize the impact of these periodic boundary conditions. As the Kaiser-Squires kernel decays to the inverse angular distance squared, the quality of the approximation due to the circular convolution quickly improves with the amount of zero-padding.

An important point to stress is that the operator $\mathbf{P}$ is unitary, with $\mathbf{P}^* \ \mathbf{P} = \mathrm{Id}$, just like the Fourier matrix $\mathbf{F}$, and as such is readily invertible. Solving \autoref{eq:forward_mass} therefore reduces to the inversion of the NDFT operator, which is the only difficult step. As mentioned in the previous section, if the spatial nodes are not regularly spaced, estimating the inverse NDFT is in the general case an ill-posed inverse problem.

Our aim is to apply the sparse regularisation framework introduced in \autoref{sec:sparse_reg} to the inversion of the NDFT operator, thus yielding an estimate of the convergence. We consider the following sparse optimisation problem:
\begin{equation}
 \bar{\bm{\kappa}} = \argmin_{\bm{\kappa}} \frac{1}{2} \parallel \Sigma^{-\frac{1}{2}}_{\gamma} \left[ \bm{\gamma} - \mathbf{T} \mathbf{P} \mathbf{F}^* \bm{\kappa} \right] \parallel_2^2 + \lambda \parallel \bm{w} \circ \bm{\Phi}^* \bm{\kappa} \parallel_1 + i_{\Im(\cdot) = 0}(\bm{\kappa}) \;.
\label{eq:conv_sparse_rec_lin}
\end{equation}
The first term is a quadratic data fidelity term, where $\Sigma_{\gamma}$ is the covariance matrix of the individual shear measurements, generally assumed to be diagonal as we are treating galaxies independently. The second term is an analysis-based sparsity constraint where $\mathbf{\Phi}$ is a dictionary providing a sparse representation of the signal we want to recover, $\circ$ is the Hadamard product (term by term product), and $\bm{w}$ is a vector of weights allowing us to adaptively adjust the $\ell_1$ ball based on the local level of noise (see \autoref{subsec:sparse_constraint}). Finally, the last term imposes the imaginary part of the solution (i.e. B-modes) to vanish. Remember that the indicator function $i_{\mathcal{C}}$ of a set $\mathcal{C}$ is defined by
\begin{equation}
i_{\mathcal{C}}(x) = \begin{cases}
	0 & \mbox{if } x \in \mathcal{C}\\
	+ \infty & \mbox{otherwise}  
\end{cases}
\end{equation}
Here, $\mathcal{C} = \left\lbrace \bm{\kappa} \in \mathbb{C}^{N \times N} \quad | \quad \Im(\bm{\kappa}) = 0 \right\rbrace$ where $\Im$ is the imaginary part, and $N \times N$ is the size of the reconstruction grid.  As a result any solution with a non-zero imaginary part is excluded. This additional constraint is crucial as the irregular galaxy sampling leads to important leakage between E- and B-modes. As E-modes are constrained by the sparsity prior, most of the signal would tend to leak towards B-modes without this additional term. Vanishing B-modes is of course a completely physically motivated prior as gravitational lensing only produces E-modes, but this also means that this method will be sensitive to spurious B-modes resulting from uncorrected systematics which will contaminate the recovered signal.

\bigskip

 We efficiently solve this problem by adopting a primal-dual algorithm following \cite{Condat2013,Vu2013}. The specialisation of this algorithm to the weak lensing mass-mapping problem is presented in \autoref{sec:appendix_prox} and summarised in \autoref{alg:2Dmassmap_lin}.
\begin{algorithm}[h]
\caption{Analysis-based $\kappa$ sparse recovery from shear}
\label{alg:2Dmassmap_lin}
\begin{algorithmic}[1]
\REQUIRE \quad \\
	Shear of each galaxy in the survey $\bm{\gamma}$.\\
	Sparsity constraint parameter $\lambda > 0$.\\
	Weights $w_i > 0$.\\
	$\tau = 2/ ( \parallel \mathbf{\Phi}\parallel^2 + \parallel \Sigma^{-\frac{1}{2}} \mathbf{T} \parallel^2)$.\\
\bigskip
\STATE $\bm{\kappa}^{(0)} = 0$
\STATE $\forall i, \quad \lambda_i^\prime = \lambda w_i$
\FOR{$n=0$ to $N_{\max}-1$}
\STATE $\nabla^{(n)} = \mathbf{F} \mathbf{P}^* \mathbf{T}^* \Sigma_\gamma^{-1} \left(\bm{\gamma} -  \mathbf{T} \mathbf{P} \mathbf{F}^{*} \bm{\kappa}^{(n)} \right)$
\STATE $\bm{\kappa}^{(n+1)} = \Re \left[ \bm{\kappa}^{(n)}  + \tau \left( \nabla^{(n)} - \mathbf{\Phi} \bm{\alpha}^{(n)} \right) \right]$
\STATE $\bm{\alpha}^{(n+1)} = \left(\mathrm{Id} - \mathrm{ST}_{\lambda^\prime}\right) \left(\bm{\alpha}^{(n)} + \mathbf{\Phi}^* \left( 2 \bm{\kappa}^{(n+1)} - \bm{\kappa}^{(n)} \right) \right) $
\ENDFOR
\RETURN $\kappa^{(N_{\max})}$.
\end{algorithmic}
\end{algorithm}

The solution of this optimisation problem tends to be biased as a well-known side effect of the $\ell_1$ sparsity constraint. In order to correct for this bias and improve the quality of the solution, we implement a two-steps approach.

We first apply the reweighted-$\ell_1$  strategy from \cite{Candes2008} which relies on iteratively solving \autoref{eq:conv_sparse_rec_lin}, adjusting each time the  weights $\bm{w}$ based on the previous estimate of the solution. This procedure is described below:
\begin{enumerate}
	\item Set the iteration count $\ell = 0$ and initialise the weights $\bm{w}^{(0)}$ according to the procedure described in \autoref{subsec:sparse_constraint}.
	
	\item Solve the weighted $\ell_1$ minimisation problem of \autoref{eq:conv_sparse_rec_lin} using \autoref{alg:2Dmassmap_lin}, yielding a solution $\bm{\kappa}^{(\ell)}$.

	\item Update the weights based on the wavelet transform of the solution $\bm{\alpha}^{(\ell)} = \Phi^t \bm{\kappa}^{(\ell)}$:
				\begin{equation}
				 w_i^{(\ell + 1)} = \left\lbrace 
				\begin{matrix} \frac{w_i^{(0)}}{|\alpha_i^{(\ell)}|/\lambda w_i^{(0)}} &\quad \mbox{ if } |\alpha_i^{(\ell)}| \geq \lambda w_i^{(0)} \\
						w_i^{(0)}  &\quad \mbox{ if } |\alpha_i^{(\ell)}| < \lambda w_i^{(0)} 
				\end{matrix}\right. \;,
			\end{equation}
	\item Terminate on convergence. Otherwise, increment $\ell$ and go to step 2.	
\end{enumerate}
In practice we find that 3 or 5 re-weightings are generally sufficient to reach a satisfying solution. However, despite improving the quality of the  solution, this reweighting scheme does not completely correct for the $\ell_1$ amplitude bias. 

The second correction step involves using this reweighted-$\ell_1$ solution to define the support  $\Omega$ of the solution in the analysis dictionary, which can then be used in place of the $\ell_1$ constraint to recover an unbiased solution. This support is defined as:
\begin{equation}
		\Omega_i = 
		\begin{cases}
		1, & \textit{if } [\mathbf{\Phi}^*\kappa^{(l)}]_i > \lambda w_i^{(0)} \\
		0, & \textit{otherwise}
		\end{cases}
\end{equation}
Using the support to constrain the solution, the recovery problem becomes:
\begin{equation}
 \bar{\bm{\kappa}} = \argmin_{\bm{\kappa}} \frac{1}{2} \parallel \Sigma^{-\frac{1}{2}}_{\gamma} \left[ \bm{\gamma} - \mathbf{T} \mathbf{P} \mathbf{F}^* \bm{\kappa} \right] \parallel_2^2 + i_{\Omega}\left( \bm{\Phi}^* \bm{\kappa} \right)  + i_{\Im(\cdot) = 0}(\bm{\kappa}) \;.
\end{equation}
where $i_{\Omega}$ is the indicator function of the  support $\Omega$. This problem preserves the sparsity of the solution of the reweighted-$\ell_1$ iterative scheme but no longer penalises the amplitude of the analysis  coefficients, which leads to an unbiased solution. This problem can still be solved using \autoref{alg:2Dmassmap_lin} simply by replacing the Soft-Thresholding operation on line 6 by  a multiplication with the support $\Omega$.

\subsection{Choice of dictionary}

\begin{figure}[t]
\centering
\begin{subfigure}[t]{0.49\columnwidth}
\includegraphics[width=\textwidth]{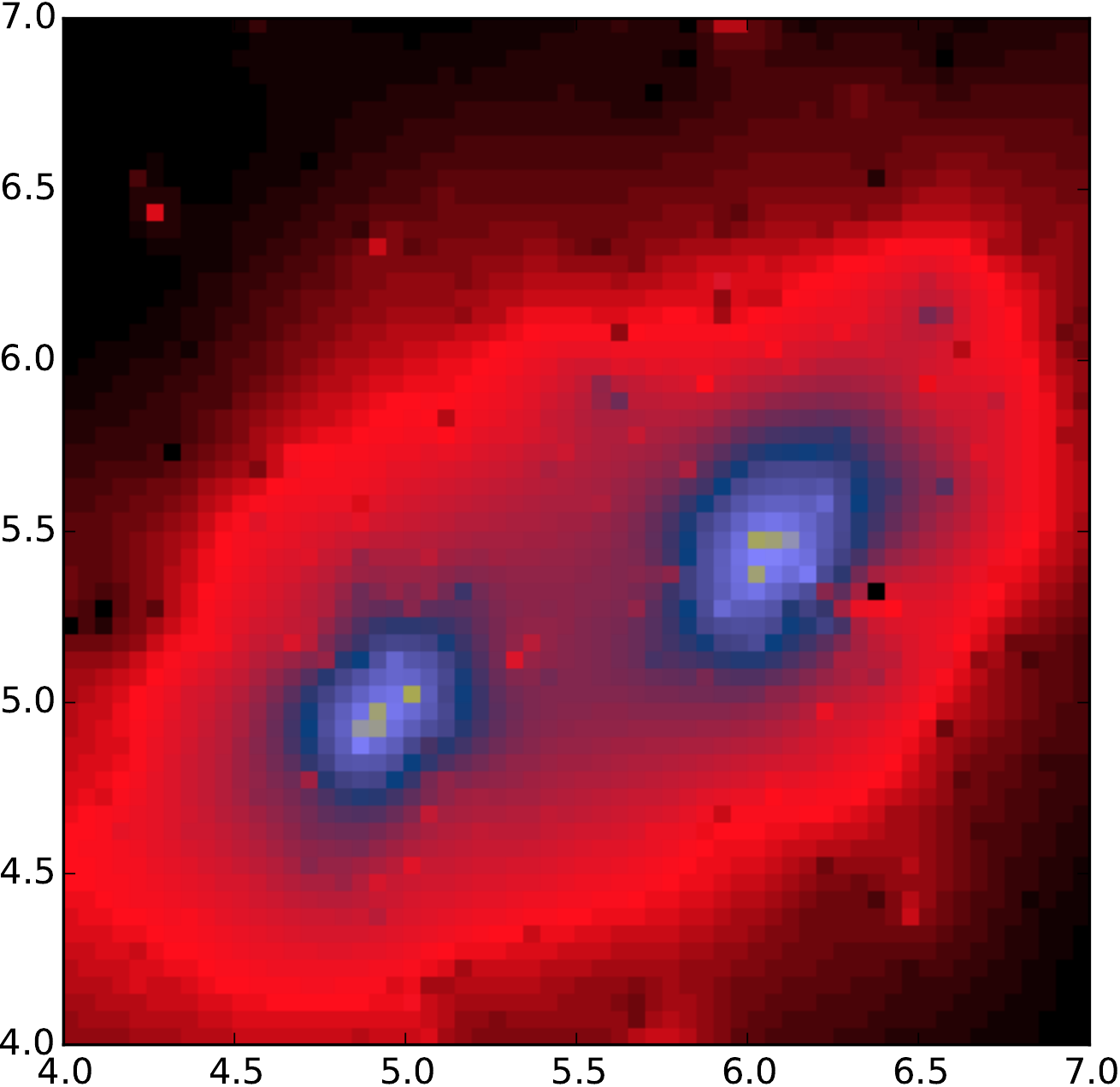}
\caption{Starlet dictionary}
\label{fig:comp_dictionary_starlet}
\end{subfigure}
\begin{subfigure}[t]{0.49\columnwidth}
\includegraphics[width=\textwidth]{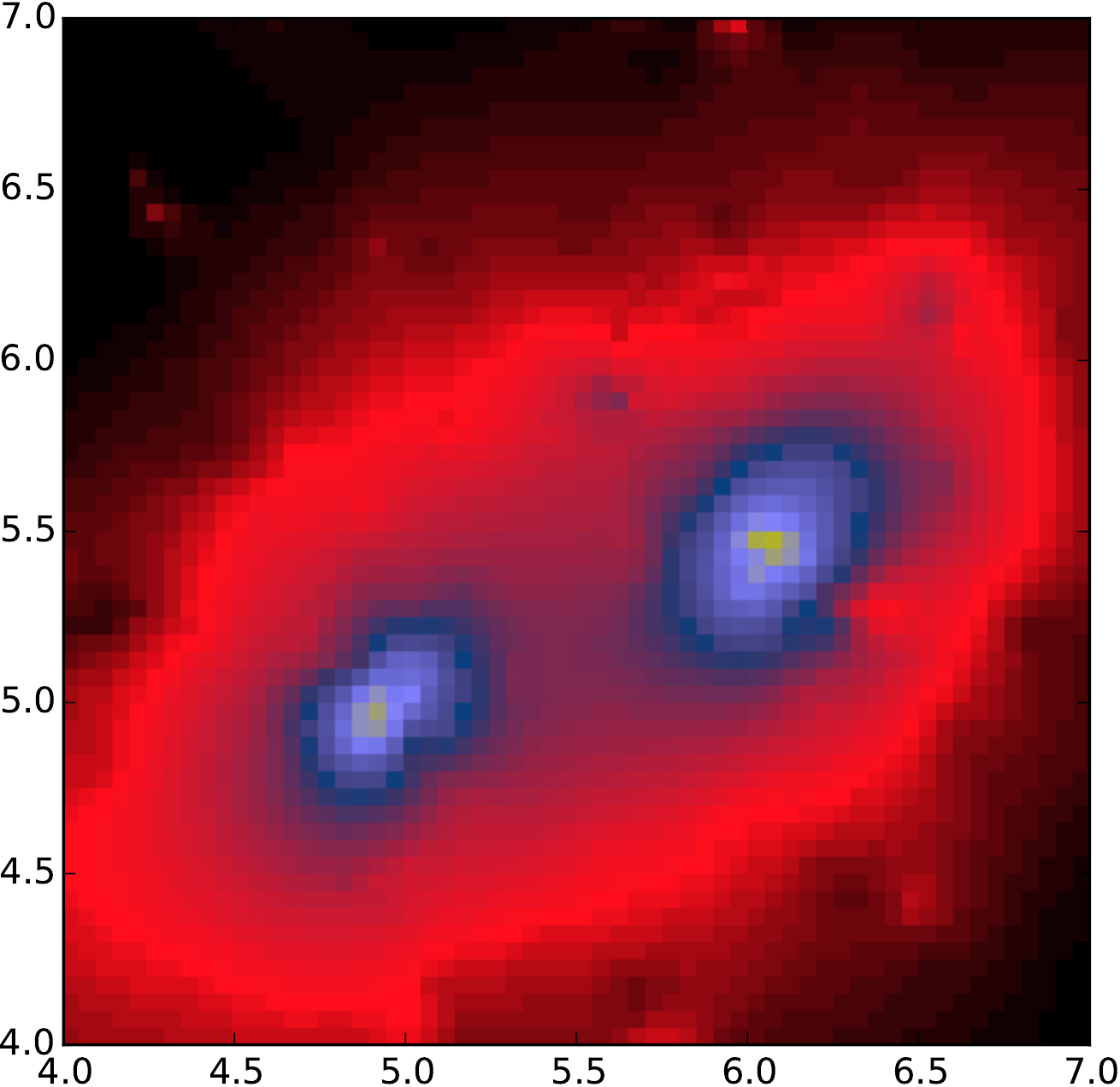}
\caption{Starlet + Battle-Lemari\'e dictionary}
\end{subfigure}
\caption{Comparison of reconstruction using starlets or a combination of starlets and Battle-Lemari\'e wavelets as the dictionary $\mathbf{\Phi}$. The Battle-Lemari\'e atoms are very effective at penalising isolated pixels.}
\label{fig:comp_dictionary}
\end{figure}

For any sparse regularisation method, an appropriate choice of dictionary is important to the quality of the result. This is especially true for noise dominated problems where the prior takes prevalence when the data is not constraining. Previous sparsity based methods developed for weak lensing mass-mapping either employed starlets for denoising \citep{Starck2006} or DCT for inpainting \citep{Pires2009}. Indeed, at small scale, the non-Gaussian convergence signal essentially generated by isolated galaxy clusters is well represented using the starlet dictionary which features isotropic atoms, adapted to the average circular profile of dark matter halos. On the other hand, on large scales, the DCT is more efficient at capturing the Gaussian part of the convergence signal and has proven to be an excellent dictionary to inpaint missing data due to masks without altering the power spectrum of the reconstructed maps.

In this work, we aim at reconstructing the convergence map on small scales and we therefore adopt the starlet dictionary for the reasons stated above. However, we find that when using the starlet alone, details at the finest scale are not sufficiently constrained, in particular spurious isolated pixels tend to contaminate the solution, as shown on \autoref{fig:comp_dictionary_starlet}. To help penalise this unwanted behaviour, we build a hybrid dictionary by concatenating to the starlet dictionary the first scale of an undecimated bi-orthogonal wavelet transform, more specifically a Battle-Lemari\'e wavelet of order 5. Contrary to starlet atoms, Battle-Lemari\'e wavelets are much more oscillatory and have a larger support (formally infinite but with an exponential decay). This makes them relatively inefficient at sparsely representing singularities such as isolated pixels, which are therefore more strongly penalised by the sparsity prior.

To illustrate the benefits of using this hybrid dictionary, we compare in \autoref{fig:comp_dictionary} the results of our reconstruction algorithm in a simple noiseless case when using starlets alone or the hybrid dictionary described above, all other parameters being kept fixed. As this simple qualitative comparison demonstrates, starlets alone tend to create small pixel-sized artefacts, even in the absence of noise. These are completely eliminated by the inclusion of the Battle-Lemari\'e wavelets. 

We add that although we find this dictionary to be effective at regularising the mass-mapping inversion, it remains generic and was not specifically designed for an optimal representation of convergence maps. More specific dictionaries could be used just as well and would potentially improve further our results. It has for instance been shown that application specific dictionaries built using Dictionary Learning  (DL) perform better than wavelets for the recovery of astronomical images \citep{Beckouche2013}.

\subsection{Adjusting the sparsity constraint}
\label{subsec:sparse_constraint}

\begin{figure*}[t]
\centering
\begin{subfigure}[t]{0.25\textwidth}
\includegraphics[width=\textwidth]{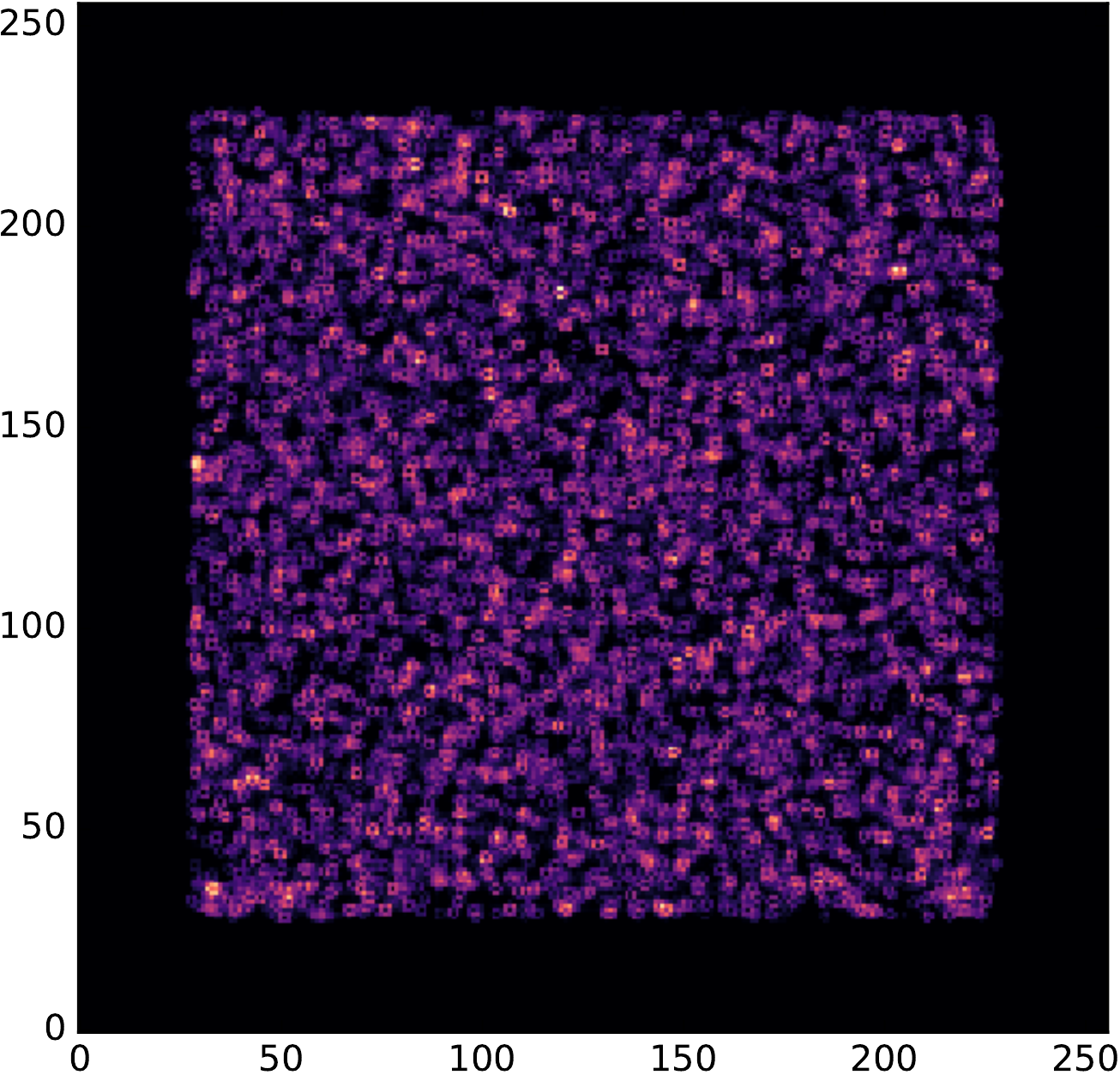}
\caption{Scale j=0}
\end{subfigure}%
\begin{subfigure}[t]{0.25\textwidth}
\includegraphics[width=\textwidth]{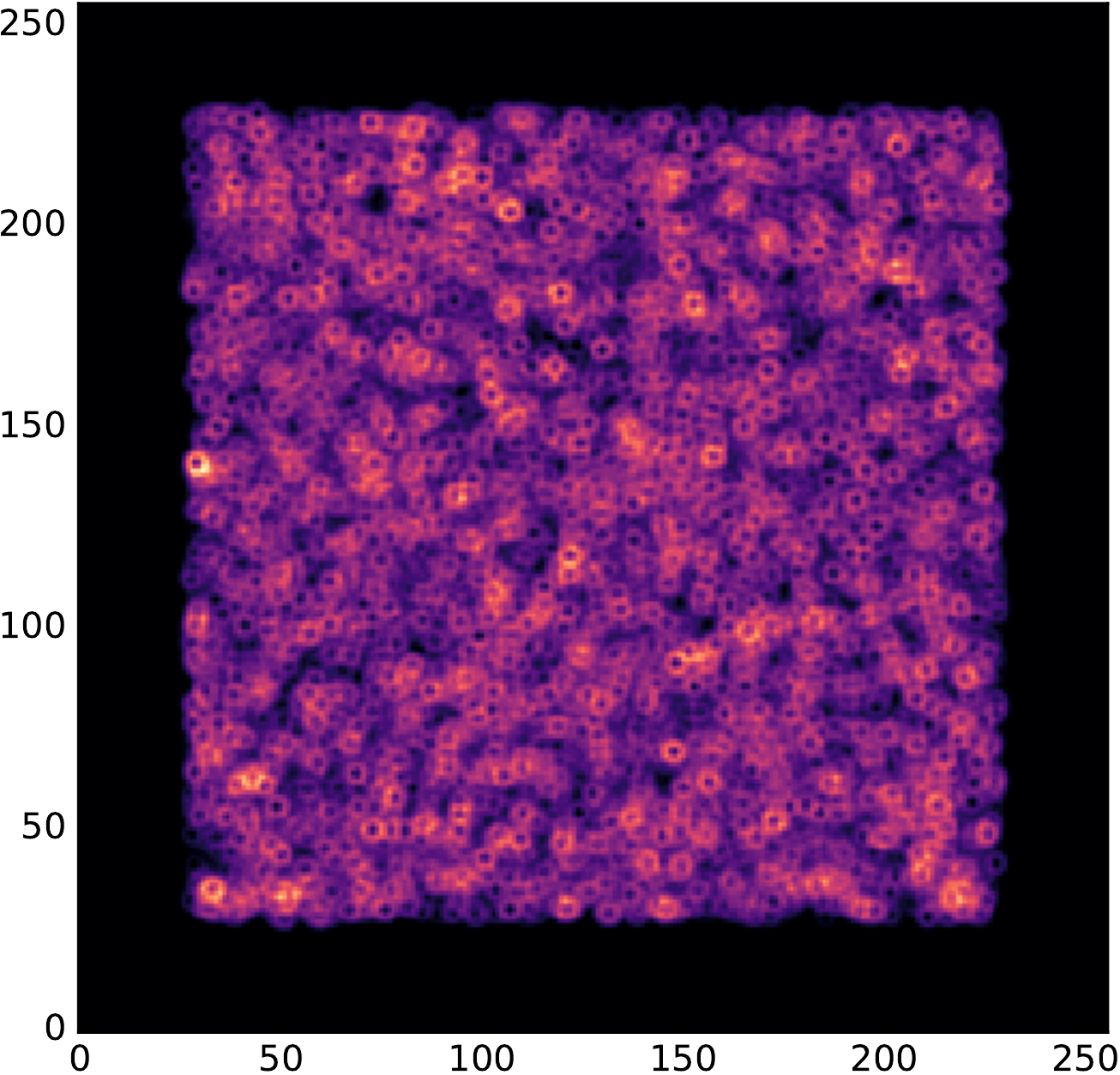}
\caption{Scale j=1}
\end{subfigure}%
\begin{subfigure}[t]{0.25\textwidth}
\includegraphics[width=\textwidth]{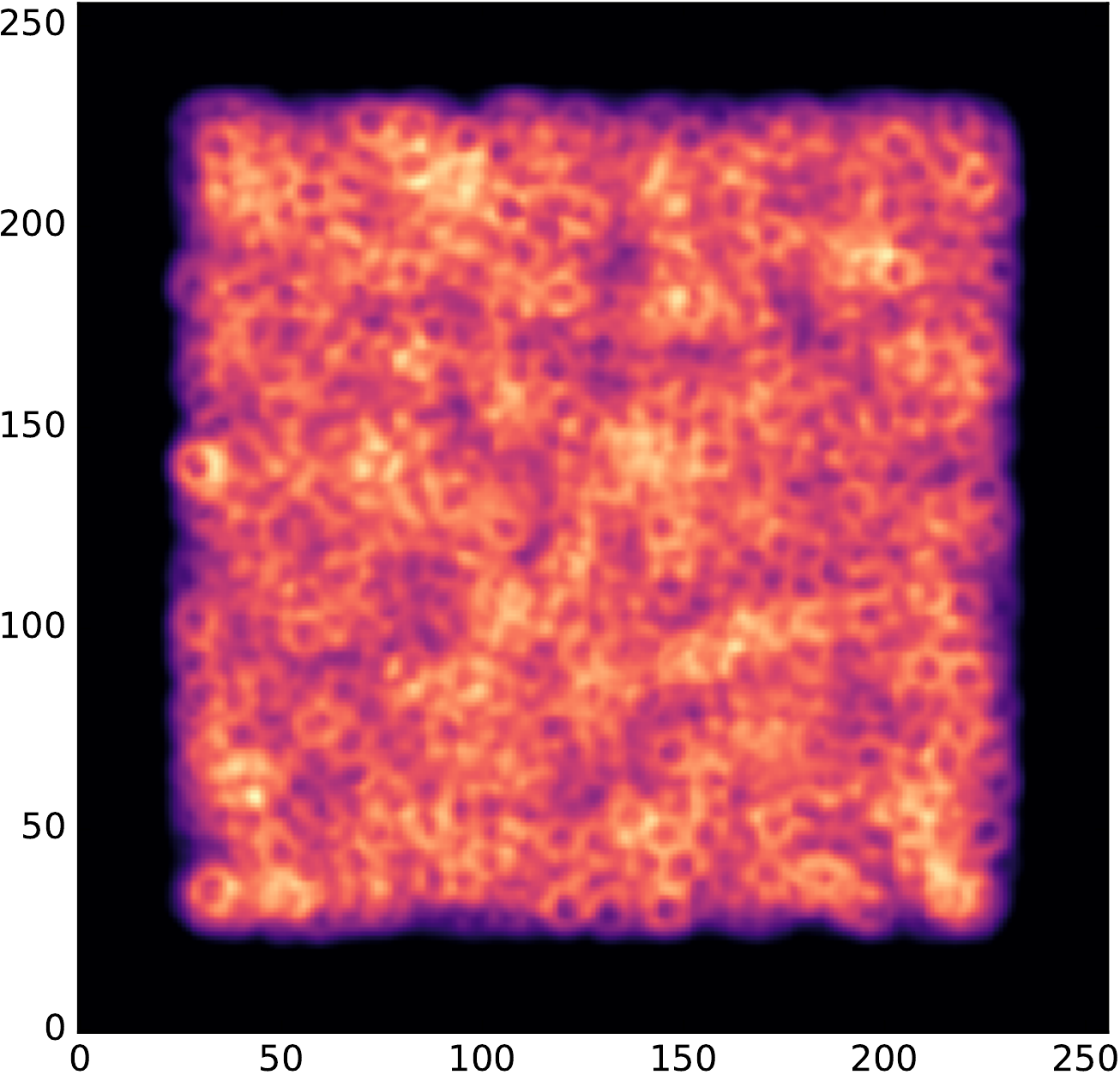}
\caption{Scale j=2}
\end{subfigure}%
\begin{subfigure}[t]{0.25\textwidth}
\includegraphics[width=\textwidth]{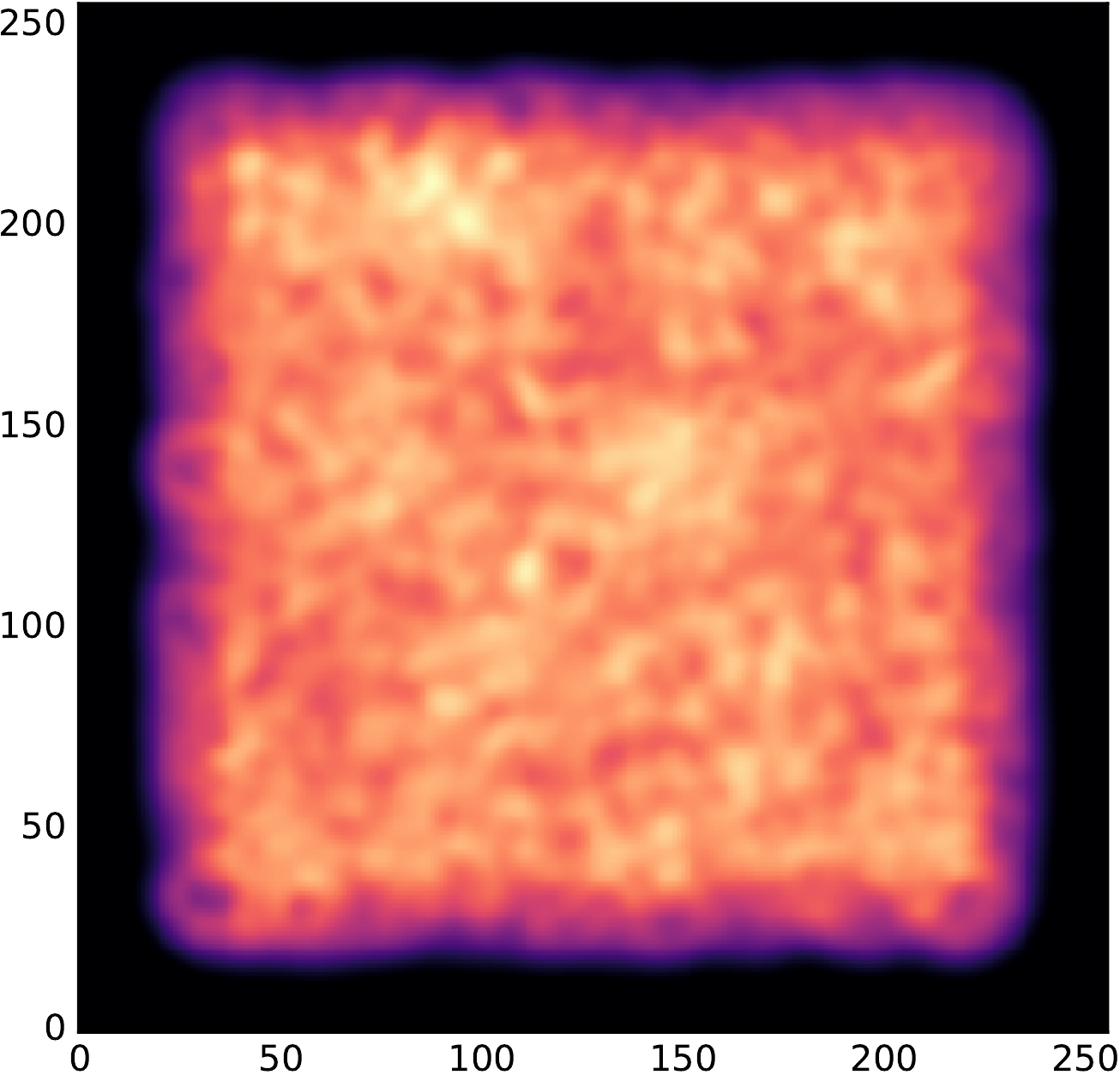}
\caption{Scale j=3}
\end{subfigure}
\caption[Standard deviation maps used to scale the sparsity constraint.]{Standard deviation maps $\bm{w}$ used to scale the sparsity constraint, obtained by propagating the shear noise to the wavelet coefficients. We show here noise maps for four successive starlet scales.}
\label{fig:noise_map}
\end{figure*}

A recurring issue with sparse recovery problems such as the one stated in \autoref{eq:conv_sparse_rec_lin} is the choice of the regularisation parameter $\lambda$. There is unfortunately no general rule indicating how to set this parameter in practice. For this application, we adopt the approach that was proposed in \cite{Paykari2014}, which consists in defining this parameter with respect to the noise level.

Formally, the parameter $\lambda$ scales the $\ell_1$ ball used in the sparsity constraint. In practice, it defines the level of Soft Thresholding applied to the dual variable $\bm{\alpha}$ (see line 6 in \autoref{alg:2Dmassmap_lin}) and therefore discriminates between significant and non-significant coefficients. While this threshold can be set according to a given sparsity model of the signal to recover, for noise dominated problems, it is much more crucial to define this threshold with respect to the noise level. As the noise statistics vary across the field, depending on the specific galaxy distribution, we introduce a vector of weights $\bm{w}$ (found in the $\ell_1$ term in \autoref{eq:conv_sparse_rec_lin}) with the purpose of locally scaling the sparsity constraint based on the standard deviation of the noise propagated to the coefficients $\bm{\alpha}$. For each wavelet coefficient $\alpha_i$ we set the weight $w_i$ to the estimated standard deviation $\sigma(\alpha_i)$. As a result, the level of 
Soft Thresholding applied to each coefficient $\alpha_i$ is $\lambda^\prime_i = \lambda w_i$ and accounts for noise variations across the field. As the threshold is proportional to the standard deviation of the noise, it can be interpreted as an hypothesis test to determine if a coefficient is due to signal or noise, assuming Gaussian statistics for the noise, which adds a powerful detection aspect to the sparsity constraint.

To estimate this noise level, and therefore set the weights $\bm{w}$, it is first necessary to understand how the noise in the data propagates to the dual variable $\bm{\alpha}$. By considering \autoref{alg:2Dmassmap_lin}, it can be seen that the noise at the level of the shear $\bm{\gamma}_N$ is propagated to the wavelet coefficients through the operation $\mathbf{\Phi}^* \Re\left(  \mathbf{F} \mathbf{P}^* \mathbf{T}^* \bm{\gamma}_N\right)$. In practice, we estimate the standard deviation of wavelet coefficients by generating Monte-Carlo noise simulations, obtained by keeping the galaxies at their observed position while randomising their orientation. Note that this step needs only to be performed once, outside of the main iteration of \autoref{alg:2Dmassmap_lin}. An example of the resulting standard deviation maps for different wavelet scales is shown on \autoref{fig:noise_map}. As can be seen, at the finest scales, the noise level has important fluctuations across the field and the contribution of 
individual galaxies can be seen. On larger scales, these local fluctuations are smoothed out and the noise level becomes much more homogeneous. 
Using this strategy therefore allows us to tune locally the sparsity constraint to take into account the specific galaxy distribution of the survey and leaves only one free parameter $\lambda$. 
\subsection{Numerical experiment on noiseless data}

The algorithm presented in this section is only meant to address the irregular sampling of the shear and the presence of noise, the complete problem being solved in the next section. In this simplified setting, we present a small numerical experiment to verify the algorithm's effectiveness at solving the linear inverse problem.

\begin{figure}[t]
\centering
\includegraphics[width=\columnwidth]{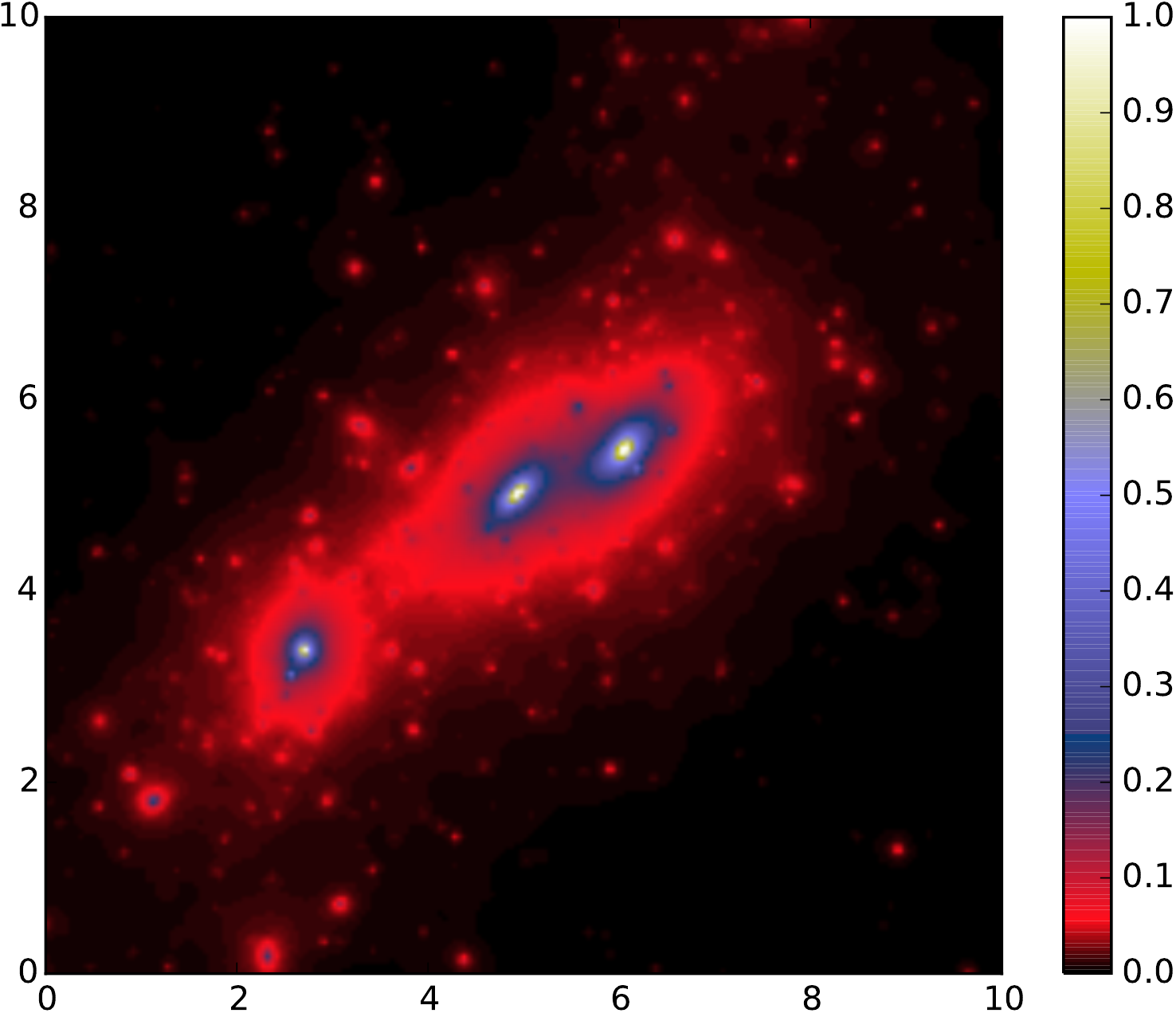}
\caption[Test convergence map]{Test convergence map generated from an N-body simulation. The clusters are located at $z_l=0.3$ while the source plane is placed at $z_s=1.2$.}
\label{fig:test_map}
\end{figure}

We simulate a $10 \times 10$ arcmin$^2$ field, containing a group of galaxy clusters extracted from the Bolshoi N-body simulations \citep{Klypin2011} (see \autoref{sec:simulations} for more details), as shown on \autoref{fig:test_map}. These clusters are placed at redshift $z_l=0.3$ and we simulate a lensing catalogue with randomly distributed sources on a single lens plane at redshift $z_s=1.2$. Note that we only simulate shear measurements from the input convergence map and not the reduced shear.

We consider here the noiseless inpainting problem where the input shear is exactly known at the position of the sources. Of course this problem is unrealistic but does illustrate the impact of missing data on the mass-mapping inversion using Fourier estimators. 

\begin{figure}[t]
\begin{subfigure}[t]{0.43\columnwidth}
\includegraphics[width=\textwidth]{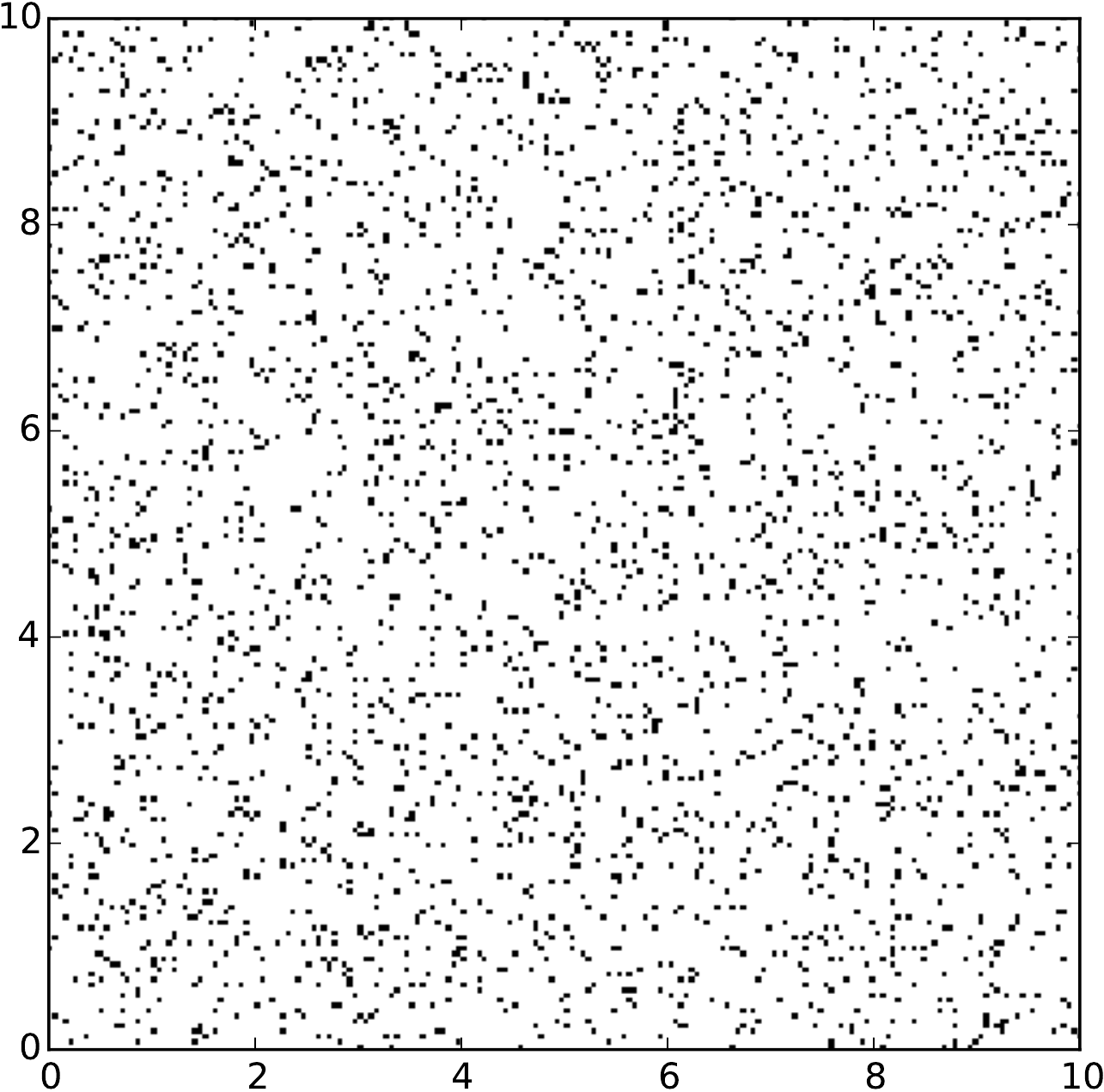}
\caption{Galaxy distribution}
\label{fig:galaxy_distribution}
\end{subfigure}\hfill
\begin{subfigure}[t]{0.49\columnwidth}
\includegraphics[width=\textwidth]{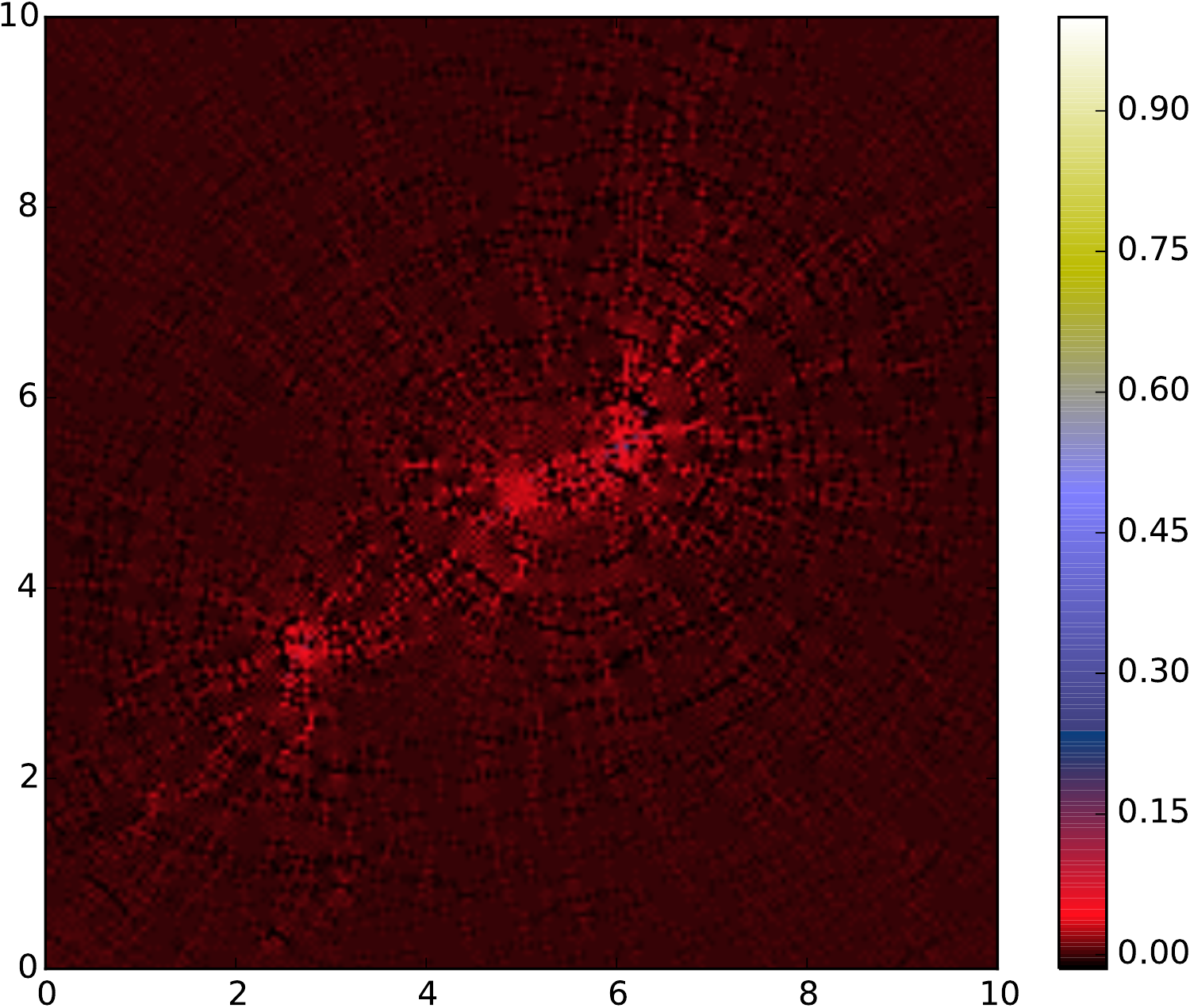}
\caption{Kaiser-Squires inversion}
\label{fig:directKS}
\end{subfigure}\\
\begin{subfigure}[t]{0.49\columnwidth}
\includegraphics[width=\textwidth]{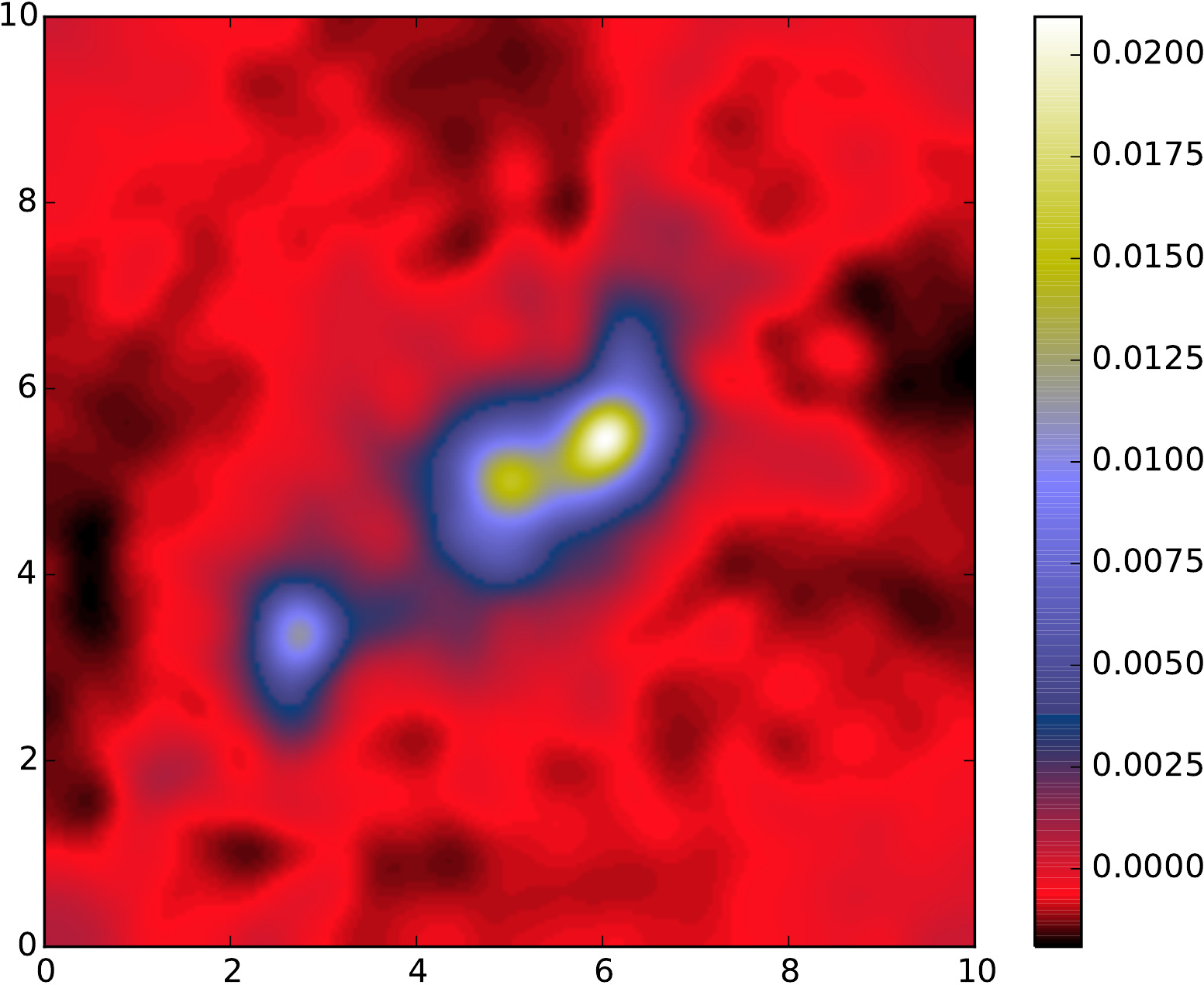}
\caption{Kaiser-Squires with a 0.25 arcmin Gaussian smoothing}
\label{fig:KSGaussian}
\end{subfigure}~
\begin{subfigure}[t]{0.49\columnwidth}
\includegraphics[width=\textwidth]{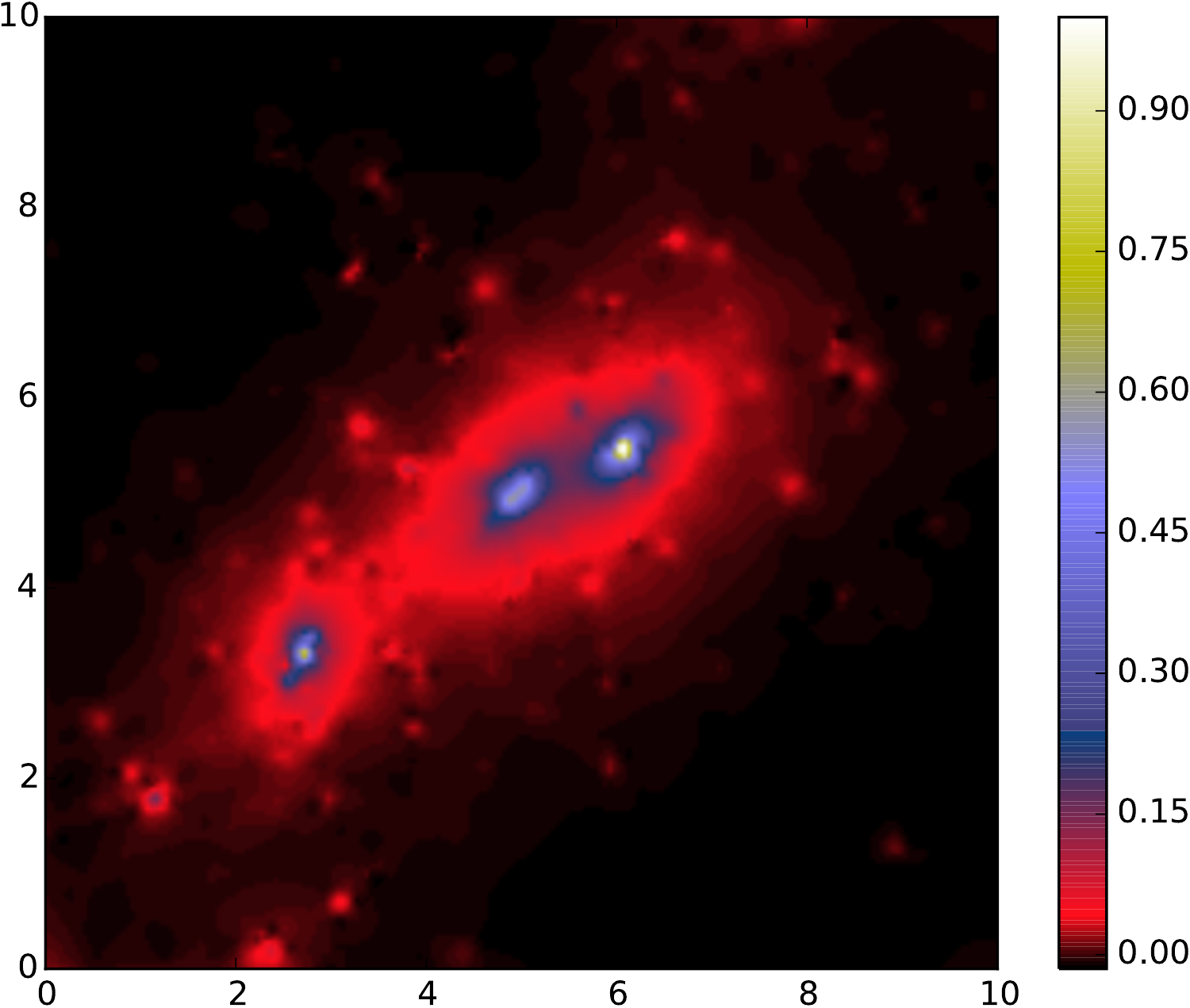}
\caption{Sparse recovery}
\label{fig:sparse_rec}
\end{subfigure}
\caption[Kaiser-Squires reconstruction from masked data.]{Reconstruction of the input convergence map using 30 galaxies per square arcminute using a Kaiser-Squires estimator (b-c) and sparse recovery (d). The top left panel shows the mask applied to the binned shear map with empty pixels marked in white. In this test, 93\% of the pixels are masked.}
\end{figure}

We generate a shear catalogue with a density of 30 galaxies per square arcminute which corresponds to a typical Euclid density. To exacerbate the effects of missing data, we reconstruct the input convergence map on a grid with 3 arcsecond pixels. Binning the input catalogue at this resolution corresponds to an average of 0.075 galaxies per pixels, which means mostly empty pixels as illustrated by the mask on \autoref{fig:galaxy_distribution} where empty pixels are shown in white. A direct Kaiser-Squires inversion with such a mask is shown on \autoref{fig:directKS} which corresponds to the solution of the inverse problem in the absence of regularisation. In order to recover the main structures we also show the result of a Gaussian smoothing with a kernel of 0.25 arcminute on \autoref{fig:KSGaussian}. Note that on this last figure the color scale is adjusted for better visualisation but the amplitude of the reconstruction is an order of magnitude below the input signal smoothed with the same 
kernel. As illustrated by these plots, small scale information is lost with this approach and although using larger bins would regularise the inversion these details would still be lost.

We then apply \autoref{alg:2Dmassmap_lin} to the same catalogue. As the weighting scheme presented in \autoref{subsec:sparse_constraint} is not meant to be used on noiseless data, we use a noisy version of the shear data to estimate the weights $\bm{w}$ and we set the regularisation parameter $\lambda$ to a very low value ($\lambda = 0.01$) to apply the algorithm on noiseless data. The solution of the optimisation problem is shown on \autoref{fig:sparse_rec}. The sparse recovery algorithm constitutes a major improvement over a simple Kaiser-Squires inversion and is able to recover pixel scale details despite 93 \% of the field being masked. Several reasons can explain these surprising results. First and foremost, the shear information is non local and even small structures formally impact the shear across all the field, albeit with an amplitude decaying to the square of the angular distance. The second reason is the use of \textit{isotropic} wavelets which provide a sparse representation of the true convergence map while being morphologically distinct from the response of individual galaxies, which is quadrupolar. As a result, the sparsity constraint greatly favours the true convergence map, which is smooth with small isotropic features, over anisotropic artefacts resulting from the irregular shear sampling.

\bigskip

The point of this numerical experiment is to show that although the shear field is randomly sampled at a relatively low rate, high frequency information can still be recovered using our method. Of course, in practice small-scale details are generally lost in the considerable amount of noise coming from intrinsic galaxy ellipticities, but not necessarily at the close vicinity of the center of galaxy clusters where the shear signal can become significant even on small scales. The ability to reconstruct small-scale details is even more relevant when including flexion information as the noise power spectrum of the flexion estimator drops at small scales.

\section{Sparse recovery for the complete problem: the Glimpse2D algorithm}
\label{sec:cluster_mapping}
In the previous section, we have introduced an algorithm based on sparse regularisation for solving the simplified linear inversion problem. Although not realistic, this problem is still an important step towards the complete surface density reconstruction that we address now in this section. We detail how the algorithm of the previous section can be modified to take into account reduced shear, redshift information for individual galaxies, and flexion information.

\subsection{Handling the reduced shear}

The problem addressed in the previous section assumed knowledge of the shear, in which case the inverse problem remains linear. While this assumption can be made in the weak regime, it breaks down at the vicinity of the structures (galaxy clusters) we are interested in mapping. Although the method presented in the previous section no longer directly applies, we present in this section how it can be extended to take into account the reduced shear $g = \frac{\gamma}{1 - \kappa}$, which makes the inversion problem non-linear.

Throughout this paper, we will restrict ourselves to the case $|\bm{g}|  \leq 1$ for simplicity, assuming that the sources which do not verify this condition can be identified and excluded from the sample. The method presented here could be extended using an iterative procedure to identify and thus treat accordingly sources lying in the region $|\bm{g}| > 1$.

\bigskip

Replacing the shear by the reduced shear in the inversion problem stated in \autoref{eq:conv_sparse_rec_lin} yields:
\begin{equation}
 \argmin_{\bm{\kappa}} \frac{1}{2} \parallel \Sigma^{-\frac{1}{2}} \left[ \bm{g} - \frac{\mathbf{T} \mathbf{P} \mathbf{F}^* \bm{\kappa}}{1 - \mathbf{T} \mathbf{F}^*\bm{\kappa}} \right] \parallel_2^2 + \lambda \parallel \bm{w} \circ \bm{\Phi}^t \bm{\kappa} \parallel_1 + i_{\Im(\cdot) = 0}(\bm{\kappa}) \;.
 \label{eq:algo_rec_non_lin}
\end{equation}
Note that at the denominator the NDFT operator is only used to evaluate the convergence at the position of each galaxy. In this form, the full problem cannot be directly addressed using the algorithm presented in the previous section as the operator to invert is no longer linear. Nonetheless, following a common strategy to handle this non-linearity, the term $\mathcal{C}_\kappa^{-1} = \Sigma^{-\frac{1}{2}}/(1 - \mathbf{T} \mathbf{F}^* \bm{\kappa})$ can be factored out and be interpreted as a diagonal covariance matrix, which depends on the signal $\bm{\kappa}$:
\begin{multline}
 \argmin_{\bm{\kappa}} \frac{1}{2} \parallel \mathcal{C}_\kappa^{-1} \left[ (1 - \mathbf{T} \mathbf{F}^* \bm{\kappa}) \bm{g} - \mathbf{T} \mathbf{P} \mathbf{F}^* \bm{\kappa} \right] \parallel_2^2 \\ + \lambda \parallel \bm{w} \circ \bm{\Phi}^t \bm{\kappa} \parallel_1  + i_{\Im(\cdot) = 0}(\bm{\kappa}) \;.
 \label{eq:algo_rec_non_lin_linearised}
\end{multline}
If the factor $\mathcal{C}_\kappa^{-1}$ is kept fixed, the problem is now linear and can be solved once again using the same class of algorithms as in the previous section. By iteratively solving this linearised problem, updating each time the matrix $\mathcal{C}_\kappa^{-1}$ with the current estimate of $\kappa$ we can recover the solution of the original problem stated in \autoref{eq:algo_rec_non_lin}. This is for instance the strategy adopted in \cite{Merten2009}.

\subsection{Including redshift information}

So far we have not given any consideration to the fact that lensing sources are not located on a single plane but are distributed in redshift. As was described in \autoref{sec:mass_sheet}, for a lens at a given redshift, the amplitude of the lensing effect experienced by each source will depend on its own redshift. Let us note $\mathbf{Z}$ the diagonal matrix of weights $Z_i$ introduced in \autoref{sec:mass_sheet}, the reduced shear can be computed from the convergence $\bm{\kappa}$ using the Fourier operators introduced thus far as:
\begin{equation}
	\bm{g} = \frac{\bm{Z} \mathbf{T} \mathbf{P} \mathbf{F}^* \bm{\kappa}}{1 - \bm{Z} \mathbf{T} \mathbf{F}^* \bm{\kappa} }
\end{equation}
where $\bm{\kappa}$ is understood to be the convergence at infinite redshift. With this new operator, the full inversion problem becomes:
\begin{multline}
 \argmin_{\bm{\kappa}} \frac{1}{2} \parallel \mathcal{C}_\kappa^{-1} \left[ (1 - \bm{Z} \mathbf{T} \mathbf{F}^* \bm{\kappa}) \bm{g} - \bm{Z}\mathbf{T} \mathbf{P} \mathbf{F}^* \bm{\kappa} \right] \parallel_2^2 \\ + \lambda \parallel \bm{w} \circ \bm{\Phi}^t \bm{\kappa} \parallel_1 + i_{\Im(\cdot) = 0}(\bm{\kappa}) \;.
 \label{eq:algo_rec_non_lin_linearised_z}
\end{multline}
where the matrix $\mathcal{C}_\kappa^{-1}$ now becomes $\mathcal{C}_\kappa^{-1} = \Sigma^{-\frac{1}{2}}/(1 -\bm{Z}  \mathbf{T} \mathbf{F}^* \bm{\kappa})$. This is simply a generalisation of \autoref{eq:algo_rec_non_lin_linearised}, which can be recovered when no redshift information is available by setting $\bm{Z} = \mathrm{Id}$, and can be solved exactly in the same way.

The main advantage of using redshift estimates for individual galaxies is the proper scaling of the resulting convergence map, which can be translated into a physical surface mass density map $\Sigma$ of the lens plane:
\begin{equation}
	\Sigma(\bm{\theta}) = \kappa(\bm{\theta})  \Sigma_{\mathrm{crit}}^{\infty} \;.
\end{equation}
The mass of the lens can then be estimated by integrating $\Sigma$ within a given radius. The second advantage of using individual redshifts is that it can help mitigate the mass-sheet degeneracy, as was described in \cite{Bradac2004} and presented in \autoref{sec:mass_sheet}. We stress however that this degeneracy cannot be completely lifted using the method presented here as, of the two terms in the gradient of the $\chi^2$ in \autoref{eq:gradient}, we only retain the one that is insensitive to the mean. Therefore, the mean value of the field remains unconstrained by the data using our algorithm. Nevertheless, we expect the additional redshift information to locally break the degeneracy and help recover the correct amplitude for the most significant structures.

\subsection{Improving angular resolution with flexion}
\label{sec:flexion}

We demonstrated in the previous section that our sparse recovery algorithm is capable of reconstructing small scale details in the absence of noise despite the irregular galaxy sampling. In practice however, the shear is noise dominated on those scales, which makes recovering high frequency information from shear measurements unlikely.

As was explained in \autoref{sec:flexion}, while the Kaiser-Squires estimator for shear has a flat noise power spectrum, the noise power spectrum of the minimum variance estimator for flexion has a $1/k^2$ dependency. As a result, although flexion measurements are generally very noisy, the noise level of the estimated map eventually drops below that of the shear on sufficiently small scales. Flexion can therefore bring useful information but only below a given scale and is very complementary to shear. 

Our aim is therefore to improve the mass-map reconstruction on small scales by extending our reconstruction method to incorporate flexion information. In the interest of simplicity, we consider here only the linear problem without redshift information, to highlight the difficulties inherent to the inclusion of flexion. The full problem is solved in the next section.

\bigskip

Following the approach developed in the first section, we address the problem using Fourier estimators. We first introduce the diagonal operator $\mathbf{Q}$ implementing the transform from convergence $\kappa$ to first flexion $\mathcal{F}$ in Fourier space, defined as:
\begin{equation}
	\hat{\bm{\mathcal{F}}} = \mathbf{Q} \hat{\bm{\kappa}} = (k_2 - i k_1) \  \hat{\bm{\kappa}} \;,
\end{equation}
where we use complex notations for the flexion with $\mathcal{F} = \mathcal{F}_1 + i \mathcal{F}_2$. Contrary to the shear operator $\mathbf{P}$, this operator $\mathbf{Q}$ is not unitary but is still invertible:
\begin{equation}
	\hat{\bm{\kappa}} = \mathbf{Q}^{-1} \hat{\bm{\mathcal{F}}} = \frac{\mathbf{Q}^*}{k^2} \hat{\bm{\mathcal{F}}} = \frac{k_2 +i k_1}{k^2} \hat{\bm{\mathcal{F}}} \;.
\end{equation}
To extend the algorithm presented in the previous section, a first straightforward approach would be to simply add a flexion term to the $\chi^2$ of \autoref{eq:conv_sparse_rec_lin} and solve the following problem:
\begin{multline}
 \argmin_{\bm{\kappa}} \frac{1}{2} \parallel \bm{\gamma} - \mathbf{T} \mathbf{P} \mathbf{F}^* \bm{\kappa} \parallel_2^2 +  \frac{1}{2}\parallel \bm{\mathcal{F}} - \mathbf{T} \mathbf{Q} \mathbf{F}^* \bm{\kappa} \parallel_2^2 \\ + \lambda \parallel \bm{w} \circ \bm{\Phi}^t \bm{\kappa} \parallel_1 + i_{\Im(\cdot) = 0}(\bm{\kappa}) \;.
\end{multline}
Although formally correct, solving this problem using the primal-dual algorithm presented in the previous section leads to a number of technical issues linked to the fact that the operator $\mathbf{Q}$, being not unitary, now contributes to the difficulty of the inverse problem. In particular, this impacts our ability to robustly identify significant coefficients in the gradient of the data fidelity term (key to the regularisation strategy presented in the previous section), which is now affected by a mixture of a flat shear noise and a flexion noise with a power spectrum in $k^2$. Furthermore, even without considering the regularisation, the inversion of the operator $\mathbf{Q}$, which is essentially a 2D gradient, requires a large number of iterations if solved with a standard gradient descent. These considerations make the algorithm much slower and far less robust to noise than when solving the problem from shear alone.

\bigskip

This needs not be the case however as the inverse of this operator is explicit in Fourier space and these difficulties can be avoided if we make proper use of this explicit inverse. We propose therefore to address the combined shear and flexion reconstruction as the following sparse optimisation problem:
\begin{multline}
	\argmin_{\bm{\kappa}, \tilde{\bm{\mathcal{F}}}} \frac{1}{2} \parallel \bm{\gamma} - \mathbf{T} \mathbf{P} \mathbf{F}^* \bm{\kappa} \parallel_2^2 +
	\frac{1}{2} \parallel \bm{\mathcal{F}} - \mathbf{T} \mathbf{F}^* \tilde{\bm{\mathcal{F}}} \parallel_2^2  \\ + \lambda \parallel \bm{w} \circ \mathbf{\Phi}^t 
	\bm{\kappa} \parallel_1 + i_{\mathrm{Im}(\mathbf{R})}\left( \left[ \begin{matrix}
	\bm{\kappa} \\
	\tilde{\bm{\mathcal{F}}}
\end{matrix}	\right]  \right)
\label{eq:conv_sparse_rec_lin_flex}
\end{multline}
where we introduce the application $\mathbf{R}: \mathbb{R}^{N \times N} \rightarrow \mathbb{C}^{2 N \times  N}, \ \bm{\kappa} \mapsto \left[ \begin{matrix}
\bm{\kappa} \\
\mathbf{F} \mathbf{Q} \mathbf{F}^*	\bm{\kappa}
\end{matrix}  \right]$, with $N \times N$ the size of the reconstruction grid.

\bigskip

Thanks to the inclusion of the auxiliary variable $\tilde{\bm{\mathcal{F}}}$ we have now decoupled the problem of the inversion of the NDFT operator $\mathbf{T}$ from the conversion between flexion and convergence which is now addressed implicitly in the last term. Remember that the indicator function $i_{\mathcal{C}}$ is infinite outside of the set $\mathcal{C}$ and therefore exclude any solution which do not belong to $\mathcal{C}$. In our case, we require the solution to be in the image of the operator $\mathbf{R}$:
\begin{equation}
	\mathrm{Im}(\mathbf{R}) = \left\lbrace \left[ \begin{matrix}
	\bm{\kappa} \\
	\tilde{\bm{\mathcal{F}}}
\end{matrix}	\right] \in \mathbb{C}^{2 N \times  N} \quad |\quad \exists \bm{\kappa} \in \mathbb{R}^{N \times N}, \quad  \tilde{\bm{\mathcal{F}}} = \mathbf{F} \mathbf{Q} \mathbf{F}^*	\bm{\kappa} \right\rbrace
\label{eq:image_of_R}
\end{equation}
The constraint $i_{\mathrm{Im}(\mathbf{R})}$ therefore implies two conditions. First, the recovered convergence has to be real, which is equivalent to enforcing the vanishing B-modes condition already used for the shear alone inversion problem. The second is that the recovered flexion $\tilde{\bm{\mathcal{F}}}$ needs to match the flexion derived from the recovered convergence, which makes the connection between the two variables $\bm{\kappa} $ and $\tilde{\bm{\mathcal{F}}}$. Interestingly, the implementation of this constraint in the primal-dual algorithm naturally gives rise to the minimum variance estimator for shear and flexion introduced in \autoref{eq:flexion_minimum_variance} and therefore will ensure that shear and flexion are optimally combined at each iteration of the algorithm to maximize the SNR of features present in the signal on all scales. 

\subsection{Complete surface density mapping algorithm}
\label{subsec:complete_dens_map}

We now present the complete reconstruction algorithm, called Glimpse2D, combining reduced shear $\bm{g}$ and reduced flexion $\bm{F}$, and taking into account individual redshift estimates for the sources. The complete problem we aim to solve is the following:
\begin{align}
 \argmin_{\bm{\kappa}, \tilde{\bm{\mathcal{F}}}} & \frac{1}{2} \parallel \mathcal{C}_{\kappa g}^{-1} \left[ (1 - \bm{Z} \mathbf{T} \mathbf{F}^* \bm{\kappa}) \bm{g} - \bm{Z}\mathbf{T} \mathbf{P} \mathbf{F}^* \bm{\kappa} \right] \parallel_2^2 \nonumber \\
 &  + \frac{1}{2} \parallel \mathcal{C}_{\kappa F}^{-1} \left[ (1 - \bm{Z} \mathbf{T} \mathbf{F}^* \bm{\kappa}) \bm{F} - \bm{Z} \mathbf{T} \mathbf{F}^* \tilde{\bm{\mathcal{F}}} \right] \parallel_2^2  \label{eq:full_conv_optim} \\
 &  + \lambda \parallel \bm{w} \circ \bm{\Phi}^t \bm{\kappa} \parallel_1 + 
  i_{\mathrm{Im}(\mathbf{R})}\left( \left[ \begin{matrix}
	\bm{\kappa} \\
	\tilde{\bm{\mathcal{F}}}
\end{matrix}	\right]  \right) \nonumber \;,
\end{align}
where the non-linear correction matrices  $ \mathcal{C}_{\kappa g}^{-1}$ and $ \mathcal{C}_{\kappa F}^{-1}$ incorporate the diagonal covariance  matrices of the shear and flexion measurement respectively.  Note that by separating shear and flexion terms, we are assuming  that these components are  independent. We use this simplistic assumption as the amplitude of the correlation between shear and flexion measurements has not been yet been quantified in practice for the AIM method. However, our formalism is trivially extensible to the full non diagonal covariance between shear and flexion, should it be provided.

As in the previous section, the  problem can be solved using the same primal-dual algorithm derived from \cite{Condat2013,Vu2013}. The specialisation of this algorithm to the full mass-mapping problem of \autoref{eq:full_conv_optim} is provided in \autoref{alg:2Dmassmap_full} and derived in \autoref{sec:appendix_full_problem}.

\begin{algorithm}[h]
\caption{Analysis-based surface density mapping algorithm from reduced shear and flexion}
\label{alg:2Dmassmap_full}
\begin{algorithmic}[1]
\REQUIRE \quad \\
	Reduced shear and flexion of each galaxy in the survey $\bm{g}$ and $\bm{F}$.\\
	Redshift weights for each galaxy in the survey $\mathbf{Z}$.\\
	Reduced shear correction matrix $\mathcal{C}_{\kappa}^{-1}$. \\
	Sparsity constraint parameter $\lambda > 0$.\\
	Weights $w_i > 0$.\\
	$\tau = 2/ ( \parallel \mathbf{\Phi}\parallel^2 + \parallel \Sigma^{-\frac{1}{2}} \mathbf{T} \parallel^2)$.\\
\bigskip
\STATE $\bm{\kappa}^{(0)} = 0$ ;  $\tilde{\bm{\mathcal{F}}}^{(0)} = 0$ 
\STATE $\forall i, \quad \lambda_i^\prime = \lambda w_i$
\FOR{$n=0$ to $N_{\max}-1$}
\STATE $\nabla^{(n)} = \mathbf{F}\mathbf{P}^* \mathbf{T}^* \mathbf{Z} \mathcal{C}_{\kappa g}^{-2} \left( (1 - \mathbf{Z} \mathbf{T} \mathbf{F}^* \bm{\kappa}^{(n)}) \bm{g} -  \mathbf{Z}\mathbf{T} \mathbf{P} \mathbf{F}^* \bm{\kappa}^{(n)} \right) + \mathbf{F} \mathbf{T}^*  \mathbf{Z}\mathcal{C}_{\kappa F}^{-2} \left( (1 -  \mathbf{Z}\mathbf{T} \mathbf{F}^* \bm{\kappa}^{(n)}) \bm{F} -  \mathbf{Z}\mathbf{T} \mathbf{F}^* \tilde{\bm{\mathcal{F}}}^{(n)} \right)$
\STATE $\left( \begin{matrix}
\bm{\kappa}^{(n+1)}\\
\tilde{\bm{\mathcal{F}}}^{(n+1)}
\end{matrix}  \right) = \prox_{\mathrm{Im}(\mathbf{R})} \left(\left( \begin{matrix}
\bm{\kappa}^{(n)}\\
\tilde{\bm{\mathcal{F}}}^{(n)}
\end{matrix}  \right)   + \tau \left( \nabla^{(n)} - \mathbf{R} \mathbf{\Phi} \bm{\alpha}^{(n)} \right) \right)$
\STATE $\bm{\alpha}^{(n+1)} = \left(\mathrm{Id} - \mathrm{ST}_{\lambda^\prime}\right) \left(\bm{\alpha}^{(n)} + \mathbf{\Phi}^t \left( 2 \bm{\kappa}^{(n+1)} - \bm{\kappa}^{(n)} \right) \right) $
\ENDFOR
\RETURN $\kappa^{(N_{\max})}$.
\end{algorithmic}
\end{algorithm}
Just as with the linear problem, we apply a reweighted-$\ell_1$ strategy to correct for the bias caused by the $\ell_1$ sparsity constraint. As the non-linear correction also requires to iteratively solve this problem we combine the update of the weights $\bm{w}$ and the update of the matrix $\mathcal{C}_\kappa$ in the following iterative procedure:
\begin{enumerate}
	\item Set the iteration count $\ell = 0$, $\mathcal{C}_\kappa^{(0)} = 1.0$ and initialise the weights $\bm{w}^{(0)}$ according to the procedure described in \autoref{subsec:sparse_constraint}.
	
	\item Solve the weighted $\ell_1$ minimisation problem of \autoref{eq:full_conv_optim} using \autoref{alg:2Dmassmap_full}, yielding a solution $\bm{\kappa}^{(\ell)}$.

	\item Update the matrices $\mathcal{C}_{\kappa g}^{(\ell)} = \Sigma^{-1}_g (1 -\bm{Z}  \mathbf{T} \mathbf{F}^* \bm{\kappa}^{(l)})$ and $\mathcal{C}_{\kappa F}^{(\ell)} = \Sigma^{-1}_F (1 -\bm{Z}  \mathbf{T} \mathbf{F}^* \bm{\kappa}^{(l)})$ 

	\item Update the weights based on the wavelet transform of the solution $\bm{\alpha}^{(\ell)} = \Phi^t \bm{\kappa}^{(\ell)}$:
				\begin{equation}
				 w_i^{(\ell + 1)} = \left\lbrace 
				\begin{matrix} \frac{w_i^{(0)}}{|\alpha_i^{(\ell)}|/\lambda w_i^{(0)}} &\quad \mbox{ if } |\alpha_i^{(\ell)}| \geq \lambda w_i^{(0)} \\
						w_i^{(0)}  &\quad \mbox{ if } |\alpha_i^{(\ell)}| < \lambda w_i^{(0)} 
				\end{matrix}\right. \;,
			\end{equation}
	\item Terminate on convergence. Otherwise, increment $\ell$ and go to step 2.	
\end{enumerate}
As for the linear problem we find that 3 to 5 iterations of this procedure are generally sufficient to reach a satisfying solution. To complement the reweighting scheme, the  same de-biasing step based on the support of the reweighted solution is applied to yield the final solution.

\section{Results on simulations}
\label{sec:sim_results}

In this section we assess the performance of the reconstruction algorithm on lensing simulations using realistic clusters extracted from N-body simulations and realistic noise levels for a space based shear and flexion survey.

\subsection{Lensing simulations}
\label{sec:simulations}
\begin{figure*}[ht!]
\includegraphics[width=\textwidth]{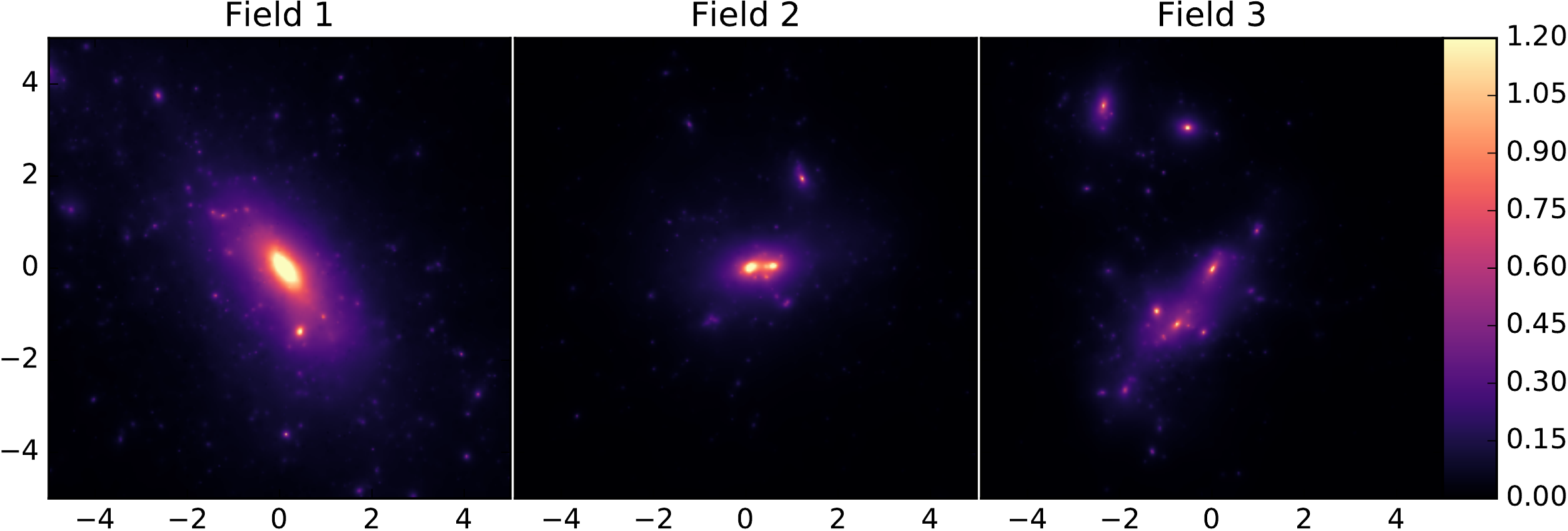}
\caption{Convergence maps for the three test clusters extracted from the N-body simulation, assuming a lens redshift of $z_L=0.3$ and background sources at infinite redshift.}
\label{fig:GT}
\end{figure*}
In order to assess the performance of the algorithm, we build a test data set based on N-body simulations. Three massive clusters were extracted at z=0 from the Bolshoi simulations \citep{Klypin2011} using the CosmoSim\footnote{\url{http://www.cosmosim.org}} interface. These clusters were selected because of their complex geometry with substantial substructure. In each case, at least one halo above $10^{13}h^{-1} $M$_{\odot}$ can be found within the virial radius of the central halo. The masses of the three clusters we consider in this work can be found in \autoref{tab:masses}.
\begin{table}[ht]
\centering
  \begin{tabular}{c c c}
  \hline \hline \\[-1.5ex]
  Field number & Virial mass  & Virial radius \\
               &   [$h^{-1} $M$_{\odot}$] &  [$h^{-1}$Mpc] \\
	\hline  \\[-1.5ex]
   1 & $1.09 \times 10^{15}$ & $2.14$ \\
   2 & $3.02 \times 10^{14}$ & $1.43$ \\
   3 & $2.70 \times 10^{14}$ & $1.45$ \\
  \hline
  \end{tabular}
  \caption{Parameters of the three halos extracted from the Bolshoi simulation}
  \label{tab:masses}
\end{table}

Density maps were obtained for these three clusters by projecting and binning the particle data with a high resolution, corresponding to pixels with an angular size of $0.5$ arcsec, for clusters located at redshift $z_L=0.3$. A multiscale Poisson denoising was subsequently applied to the binned density maps to remove the shot noise from the numerical simulation. The resulting surface mass density maps were then scaled to create convergence maps corresponding to clusters at a redshift $z_L=0.3$ lensing background galaxies at infinite redshift. Although we acknowledge that these three particular clusters, selected at $z=0$, are not necessarily representative of clusters at $z_L=0.3$, for the purpose of testing the mapping algorithm they provide more realistic density distributions than simple models based on SIS or NFW profiles. Shear and flexion maps were then derived from these reference convergence maps and only the center $10 \times 10$ arcmin central region of each field is kept. At the redshift of the lenses, this corresponds to a physical size of $1.88 \times 1.88 ~ h^{-1}$Mpc. \autoref{fig:GT} illustrates the convergence maps scaled for sources at infinite redshift for the 3 fields we consider.

Finally, 100 mock galaxy catalogues were produced for the three fields using for each realisation a uniform spatial distribution of background galaxies with a density of 80 gal/arcmin$^2$ and the following redshift distribution:
\begin{equation}
p(z) \propto \frac{z^2}{2 z_0^3} \exp(-z /z_0)
\end{equation}
with $z_0=2/3$. The median redshift of this distribution is $z_{med} = 1.75$ and we truncate the distribution at $z_{max} = 5$. This particular distribution has been used in a number of different works and in particular in \cite{Cain2015} to represent the actual galaxy distribution for a typical HST/ACS field. We compute the reduced shear and flexion for each source galaxy based on their redshift and on the resdshift of the lens. In the final mock catalogues we assume Gaussian photometric redshifts errors for each sources with $\sigma_z = 0.05 (1 + z)$, an intrinsic shape noise of  $\sigma_\epsilon = 0.3$ for the reduced shear measurements and $\sigma_F = 0.029$ arcsec$^{-1}$ for the reduced flexion measurements. This particular value for the flexion noise is in accordance with previous works \citep{Rowe2013} and corresponds to the median dispersion of the flexion measurements obtained using the
AIM method on HST data for the Abel 1689 cluster \citep{Cain2011}.

Some final cuts are applied to the galaxy catalog to exclude strongly lensed sources by discarding all sources which verify $|g| \geq 1$ or $|F|  \geq 1.0$ arcsec$^{-1}$. The flexion cut is based on recommendations from \cite{Cain2015} and constitutes a simple approximation to the practical limit encountered in real data when estimating flexion for extremely lensed sources.

This particular setting was chosen to match the simulations used in \cite{Cain2015} so that our results can be qualitatively compared to theirs. However, we stress that contrary to their work, our reconstruction method relies only on shear and flexion and does not include strong lensing constraints which are very powerful to map the very center of the clusters.

Finally, throughout this section, we assume a fiducial $\Lambda$CDM model for computing distances with $\Omega_m=0.25$, $\Omega_\Lambda =0.75$ and $H_0 = 70$ km/s/Mpc. Cosmological computations are implemented using the NICAEA library\footnote{\url{http://www.cosmostat.org/software/nicaea}}.

\subsection{Results}

For each of the mock catalogues generated, we perform the inversion with and without the flexion information to assess how much the flexion information can help recover the substructure of the clusters. The same parameters are used for all the three different fields and are summarised in \autoref{tab:parameters}. The pixel size can in theory be arbitrarily small but we choose 0.05 arcmins as a good compromise between resolution and computational cost. The number of wavelet scales is not crucial to the quality of the reconstruction, we adjust it so that the maximum scale corresponds roughly to the order of the maximum size of the structures we want to recover. Finally, the number of iterations and reweighting steps is not crucial either as the algorithm does converge to a solution.
\begin{table}[h]
\centering
  \begin{tabular}{l c}
  \hline\hline \\[-1.5ex]
  Parameter & Value \\
  \hline \\[-1.5ex]
  Pixel size & 0.05 arcmin \\
  Number of wavelet scales & 7 \\
  $K-$sigma threshold & 5 \\
  Number of iterations & 500 \\
  Number of re-weightings & 5\\
  \hline
  \end{tabular}
  \caption{Parameters of the reconstruction algorithm}
  \label{tab:parameters}
\end{table}

Assessing the quality of the inversion from a single realisation of the galaxy catalog is difficult as the ability to detect small structure will depend on the specific positions of the galaxies in one realisation. Therefore, we compute the mean and standard deviation of the reconstructed maps over 100 realisations of both shape noise and galaxy positions. 
\begin{figure*}
\centering
\includegraphics[width=\textwidth]{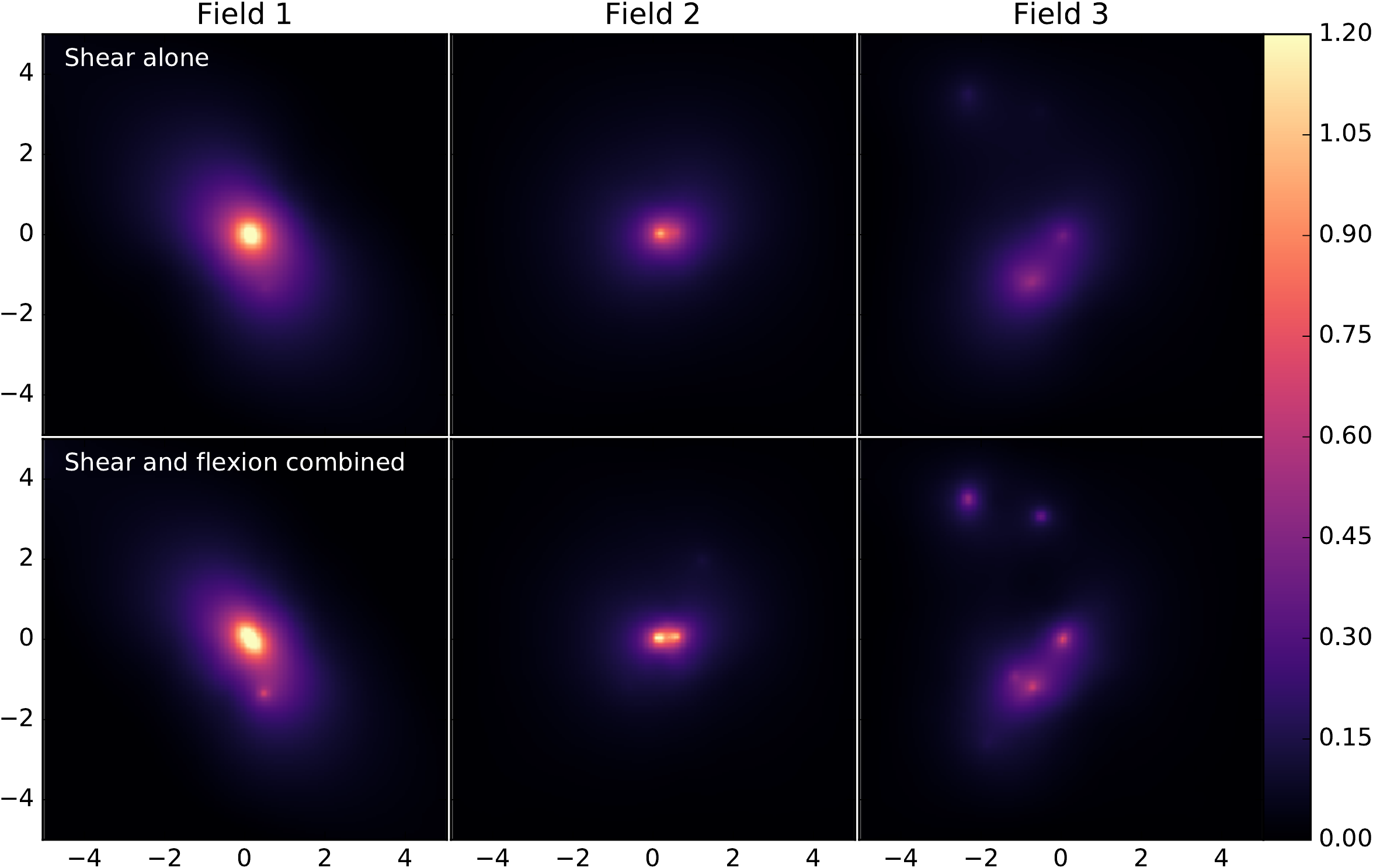}
\caption{Mean of 100 independent signal realisations for the three fields. The top row corresponds to reconstructions using the shear alone while the bottom row corresponds to reconstructions using shear and flexion.}
\label{fig:mean}
\end{figure*}
The mean of the reconstructions for the 3 fields with and without flexion is represented in \autoref{fig:mean}. The most striking difference between the 2 sets of reconstructions is the flexion's ability to detect very small substructures at the 10 arcseconds scale which are not detected from shear alone. This is clearly visible for the first and third fields.

\begin{figure}
\includegraphics[width=\columnwidth]{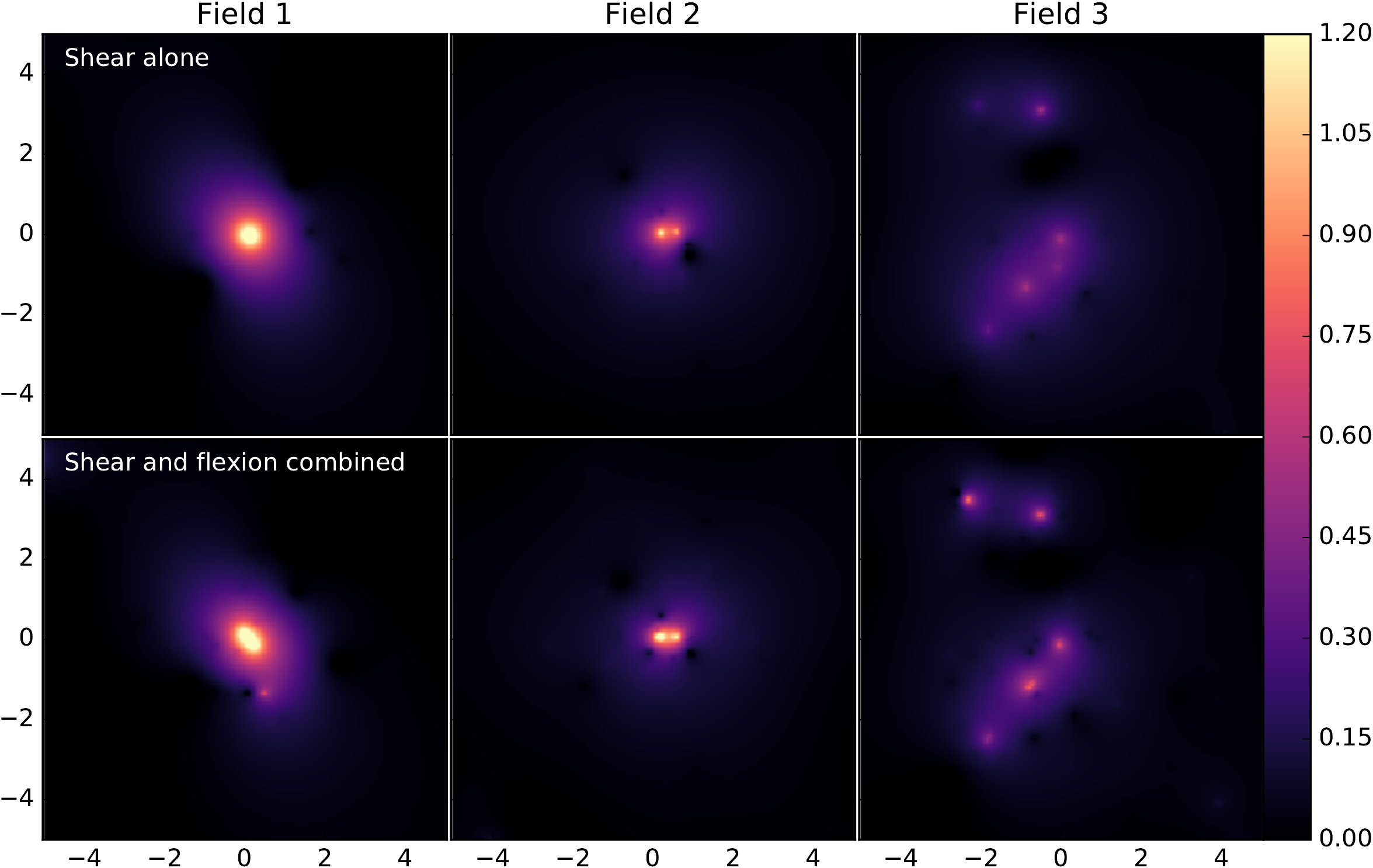}
\caption{Reconstruction of the 3 clusters for a single noise realisation. The top row corresponds to reconstructions using the shear alone. The bottom row corresponds to simulations using shear and flexion.}
\label{fig:rec0}
\end{figure}

\begin{figure}
\includegraphics[width=\columnwidth]{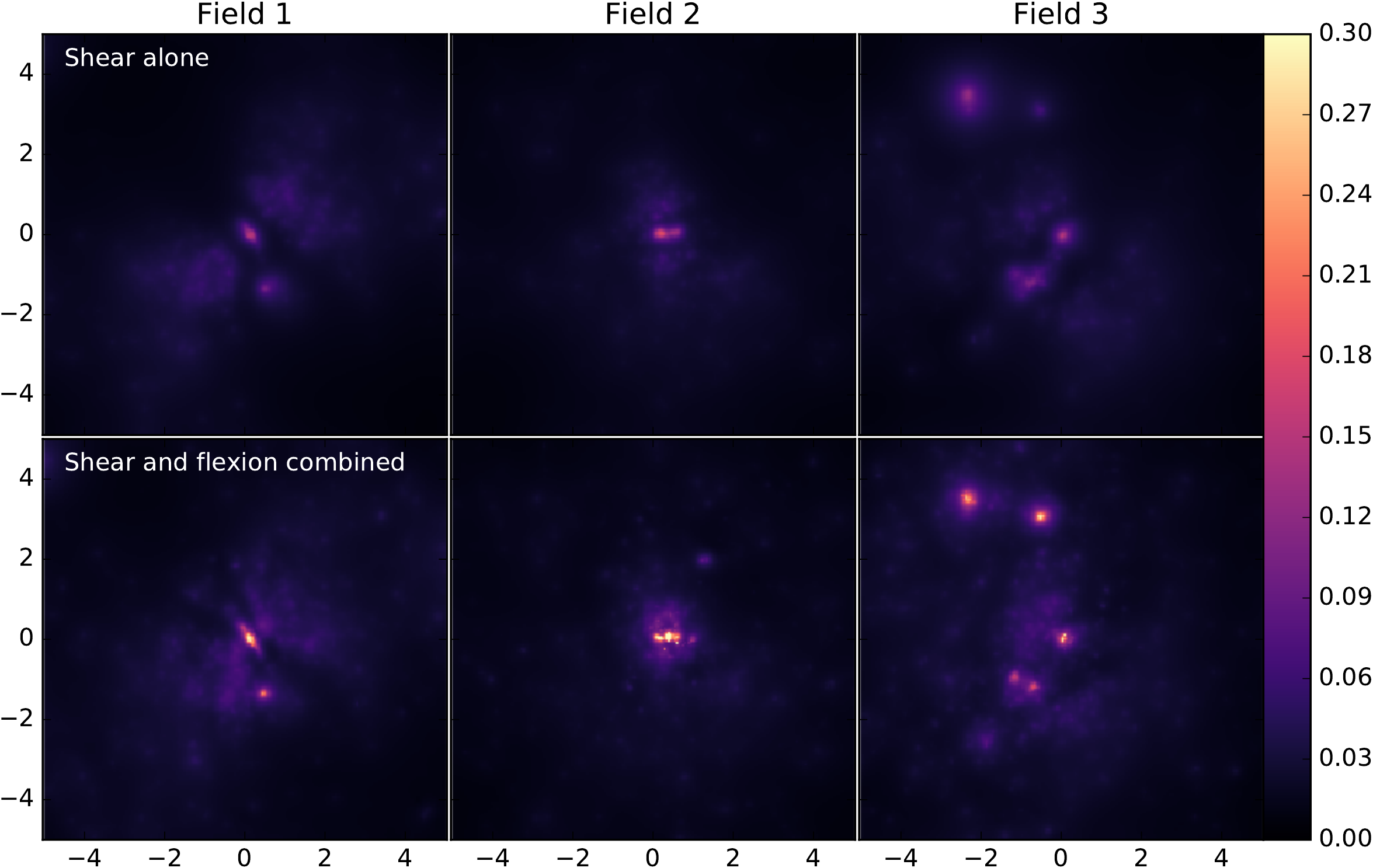}
\caption{Standard deviation of the 100 independent signal realisations for the 3 clusters. The top row corresponds to reconstructions using the shear alone. The bottom row corresponds to simulations using shear and flexion.}
\label{fig:std}
\end{figure}

The errors on the reconstructed mass maps are estimated by taking the standard deviation of each pixel in the 100 independent mock catalogues realisation. It is important to stress that the observed dispersion of the reconstructions are due to 2 effects: different noise realisations and different galaxy distributions in angular position and redshift. The standard deviation with and without flexion for all 3 fields is illustrated in \autoref{fig:std}. As can be seen on this figure, we only observe significant dispersion on small scales, around detected structures. This is expected as these scales have the lowest signal to noise ratio in the data which makes the reconstruction of these features strongly dependent on the specific noise and galaxy distribution realisation. Nevertheless, we find that the reconstruction is remarkably stable between realisations. \autoref{fig:rec0} shows reconstructions of the mass maps for one realisation of the galaxy catalogues. We see that for a single realisation the reconstructed maps is very close to the mean maps and mainly differs for small substructure. For instance, the sub-halo at the bottom of field 3 appears stronger in this particular realisation than in the average maps. Such deviations are unavoidable but the shape of the halos can reliably be reconstructed from a single data set, as would be the case for actual data.

Flexion does not only enable the detection of very small structures, it also helps to constrain the small scale shape of the main halos. Indeed, in shear reconstructions, the noise dominates the signal at these scales which makes the sparsity prior apparent on small scales. As we are using isotropic wavelets, without sufficient evidence from the data, the reconstruction will be biased towards isotropic shape. This is visible for instance at the center of the first halo, on the top left image of \autoref{fig:mean}, where the elongation of the very center of the cluster, visible in \autoref{fig:GT}, is clearly lost. On the contrary, including flexion information helps to constrain the shape of the center of the cluster. 

\begin{figure*}
\centering
\includegraphics[width=\textwidth]{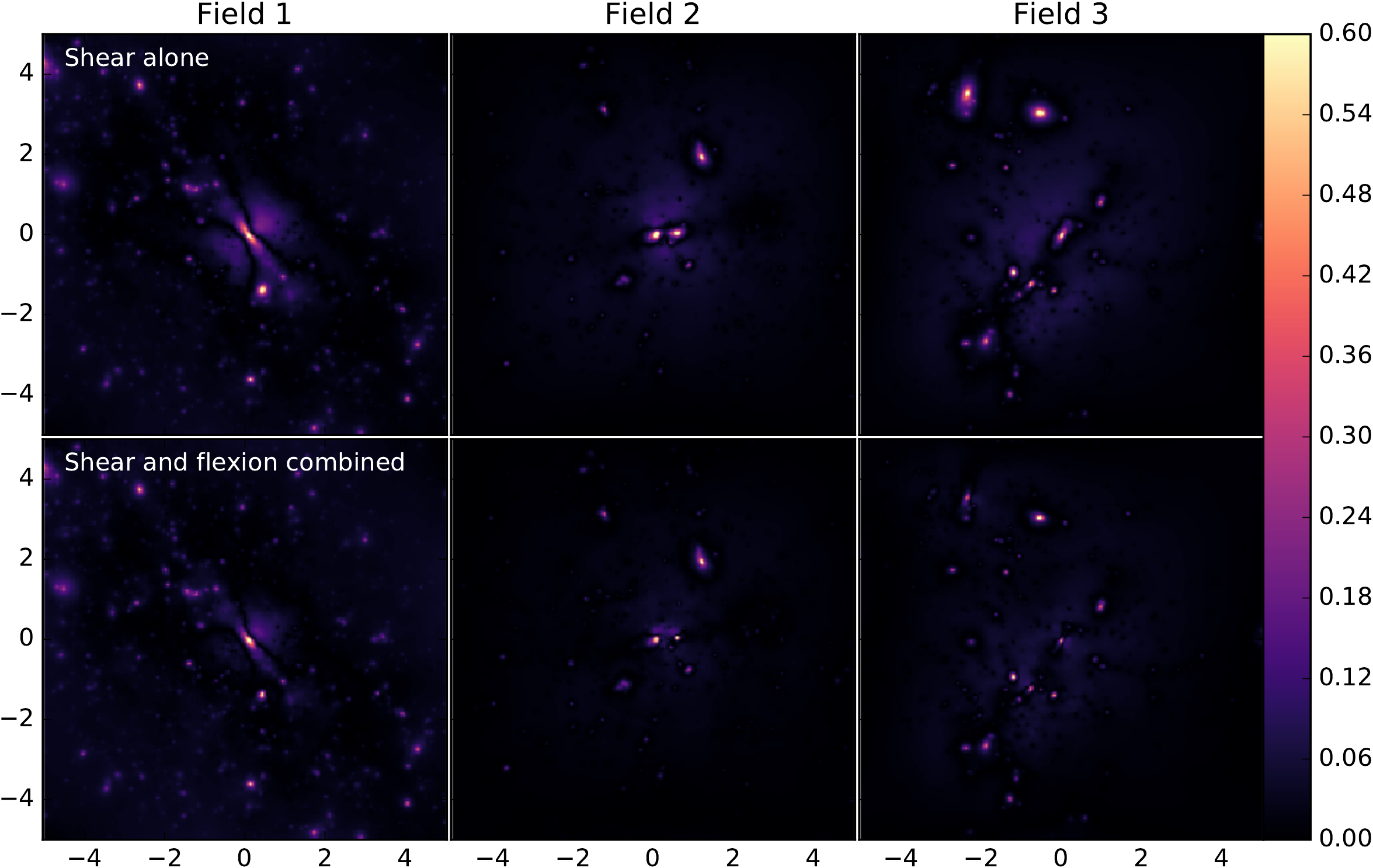}
\caption{Absolute difference between the mean  of 100 independent signal realisations and the input convergence maps. The top row corresponds to reconstructions using the shear alone while the bottom row corresponds to reconstructions using shear and flexion.}
\label{fig:diff}
\end{figure*}

The same effect can be seen in \autoref{fig:diff} where we show the absolute difference between the mean of the reconstructions and input convergence maps. The clusters are successfully recovered on large scales in both cases, leaving mostly only the very fine substructure visible in these residual maps. Nevertheless, although small, some residuals of the main halos remain visible near the center of  the  center fields but are lower when combining shear and flexion, indicating better recovery of the  shape of the central part of the cluster when flexion is included. This is particularly visible for the first cluster where the shear only residuals clearly indicate that the anisotropy of the halo is not captured as well by the shear alone reconstruction than by the combined shear and flexion reconstruction.

We also note that on large scales, where the shear dominates, the two sets of reconstructions agree with each other and the inclusion of flexion does not modify the reconstruction. This is illustrated by \autoref{tab:rec_masses} where the integrated mass within a $1^{\prime}$ radius is computed on the reconstructions with and without flexion. As can be seen, tight constraints on the halo masses can be obtained for the three fields considered from shear only and the addition of flexion does not change the mass estimates on scales larger than the arcminute.
\begin{table}[ht]
\centering
  \begin{tabular}{c c c c}
  \hline \hline\\[-1.5ex]
  Field  & Input $M_{1^\prime}$  & Shear only $M_{1^\prime}$ & Combined $M_{1^\prime}$ \\
           &   $10^{13} ~h^{-1} $M$_{\odot}$ & $10^{13} ~h^{-1} $M$_{\odot}$ &  $10^{13} ~h^{-1} $M$_{\odot}$\\ 
   \hline\\[-1.5ex]
   1 & $18.5$ & $19.1 \pm 0.53$ & $18.7 \pm 0.54$ \\
   2 & $9.48$ & $10.5 \pm 0.62$ & $10.2 \pm 0.62$ \\
   3 & $5.88$ & $6.89 \pm 0.83$ & $6.97 \pm 0.71$ \\
  \hline
  \end{tabular}
  \caption{Integrated mass in the central 1$^{\prime}$ of each field, corresponding to a radius of $ 188 \ h^{-1} $kpc at the redshift of the lens.}
  \label{tab:rec_masses}
\end{table}

\begin{figure*}
\includegraphics[width=\textwidth]{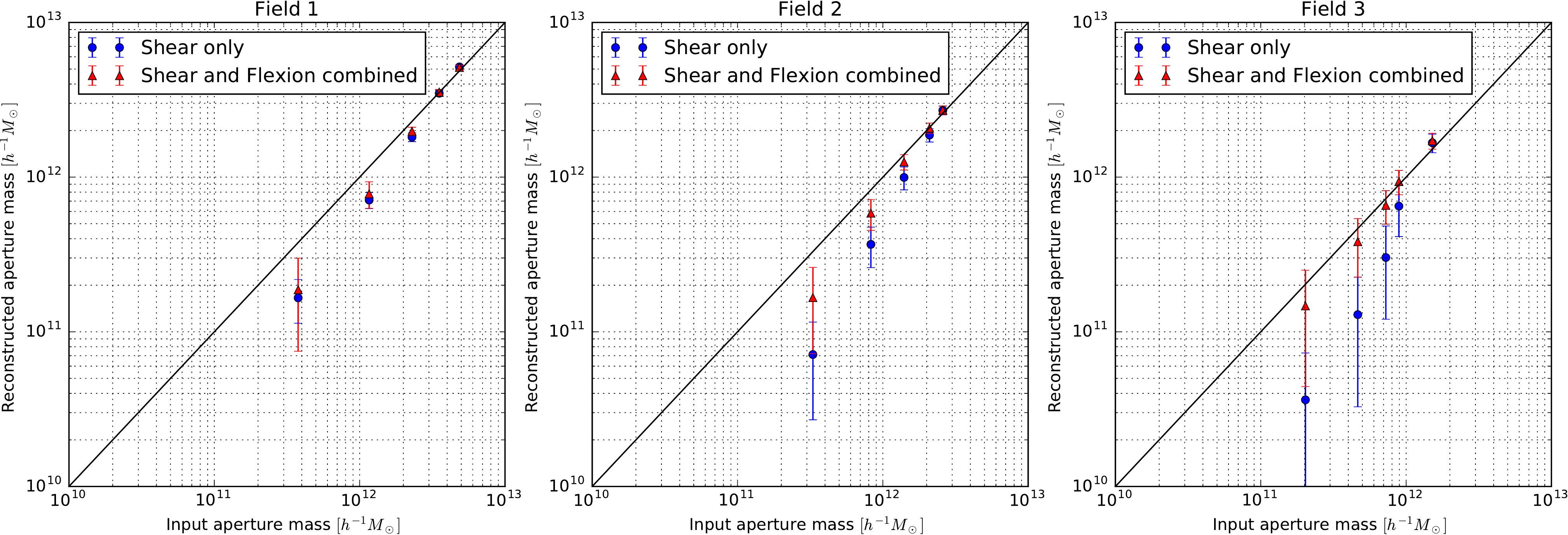}
\caption{Aperture masses measured at the centre on the main halo for each field and for five different radii: 19, 38, 75, 150, and 300 $h^{-1}$ kpc or 6$^{\prime\prime}$,12$^{\prime\prime}$,24$^{\prime\prime}$,  48$^{\prime\prime}$, and 1.6$^\prime$. The blue circles indicate the masses computed on the shear only reconstructions while the red triangles indicate the masses computed on the combined shear and flexion reconstructions. Error bars represent the standard deviation of 100 independent realisations.}
\label{fig:centerSubMass}
\end{figure*}

To quantify the impact of flexion on the recovery of small-scale  structures, we show in \autoref{fig:centerSubMass} aperture masses for five different radii (19, 38, 75,  150, and 300 $h^{-1}$ kpc or 6$^{\prime\prime}$,12$^{\prime\prime}$,24$^{\prime\prime}$,  48$^{\prime\prime}$, 1.6$^\prime$) at the  center of the 3 fields and compare them with the input aperture masses. As can be seen for all three fields, the aperture masses computed for the largest radius, which is above the arcminute scale coincide between the shear only and combined shear and flexion reconstruction, showing that on large scales the reconstructions are mostly constrained by the shear information. Furthermore, above the arcminute scale, the measured aperture masses are in excellent agreement with the input masses for all three fields. The  impact of flexion becomes obvious on small scales where one can see that the aperture masses are systematically underestimated from the shear only reconstruction while adding flexion can dramatically improve the recovery of small scale details as can be seen for field 3.

\begin{figure}
\includegraphics[width=\columnwidth]{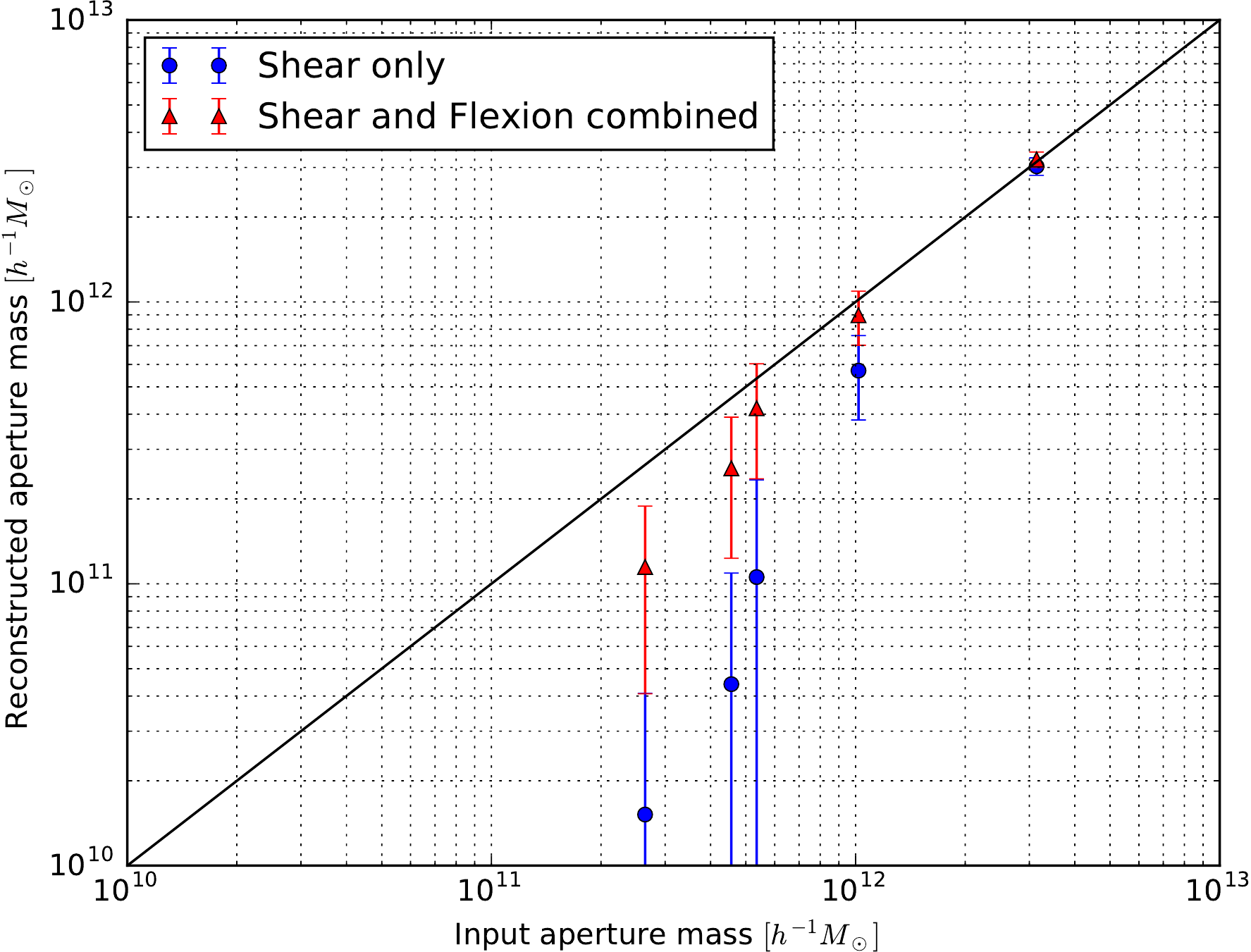}
\caption{Aperture mass for the main subhalo in field 1 for five different radii: 19, 38, 75, 150, and 300 $h^{-1}$ kpc or 6$^{\prime\prime}$,12$^{\prime\prime}$,24$^{\prime\prime}$,  48$^{\prime\prime}$, and 1.6$^\prime$. The blue circles indicate the masses computed on the shear only reconstructions while the red triangles indicate the masses computed on the combined shear and flexion reconstructions. Error bars represent the standard deviation of 100 independent realisations.}
\label{fig:submass}
\end{figure}

To  illustrate further the impact of flexion on the recovery of substructure, we also show in \autoref{fig:submass} the aperture mass at the same five radii but for  the subhalo visible in the combined shear and flexion reconstruction of  field 1.  As can be seen in this figure, this substructure, which is around the 15$^{\prime\prime}$ scale, is clearly recovered by the addition of flexion between the 12$^{\prime\prime}$ and 24$^{\prime\prime}$ scale.

\section{Conclusion}

In this work, we have introduced a new mass-mapping methodology, aimed at the recovery of high-resolution convergence maps from a combination of shear and flexion measurements. In order to preserve all available small-scale information, our approach avoids any binning or smoothing of the input lensing signal and still relies on efficient Fourier estimators despite the irregular  sampling of the shear and flexion fields. Stating the mass-mapping problem using Fourier estimators on an irregularly sampled  lensing field is an ill-posed linear inverse problem, which we solve using the sparse regularisation framework. Our full reconstruction algorithm incorporates individual redshift, shear and flexion measurements for background galaxies and accounts for the non-linear reduced shear and flexion. We define our sparse regularisation based on a multiscale noise variance map derived from randomisations of the data which allows us to tune the regularisation by a single parameter corresponding to a significance level for the detection of structures.

We demonstrate the effectiveness of this approach for dealing with irregularly sampled data in the simplified linear case of reconstructing the convergence from noiseless shear measurement. We show near-perfect recovery of an input convergence field despite 93\% of missing data,  at a 3$^{\prime\prime}$ resolution with only 30 background galaxies per arcmin$^2$.

In order to test our reconstruction algorithm in a realistic setting and, more  importantly, to demonstrate the added value of flexion, we build a set of mock catalogues corresponding to a deep HST/ACS survey using as lenses realistic dark matter distributions extracted from N-Body simulations. We compare the reconstructions of these simulated fields with and without the inclusion of flexion. 

We verify that, the shear alone and combined shear and flexion reconstructions agree on scales larger than the arcminute. This behaviour is expected, by construction, as our algorithm weighs the shear and flexion information based on their respective noise levels, promoting shear on large scales and flexion on small scales. More interestingly, we demonstrate that the inclusion of flexion impacts the recovery of sub-arcminute scales in two different ways. First and foremost, substructures at the 15$^{\prime\prime}$ scale which are lost when reconstructing from shear alone can be recovered by the addition of flexion. Second, the inclusion of flexion allows a much better constraint of the shape of the inner part of the main halos, in particular we find that anisotropies are more efficiently recovered. These results are in good agreement with the conclusions of \cite{Cain2015}, showing the importance of flexion for the recovery of high resolution maps well outside of the inner-most regions of clusters where strong lensing prevails.

In the spirit of reproducible research, our C++ implementation of the mass-mapping algorithm presented in this paper will be made freely available at:
\begin{center}
	\url{http://www.cosmostat.org/software/glimpse}
\end{center}
The mock catalogues and their corresponding reconstructions as well as the input convergence maps are available at \url{http://www.cosmostat.org/product/benchmark_wl}.

\begin{acknowledgements}
The authors thank Jalal Fadili, Eric Jullo, Fred Ngole, Chieh-An Lin and Martin Kilbinger for useful comments and discussions. This work was funded by the Centre Nation d’Etude Spatiale (CNES) and the DEDALE project, contract no. 665044, within the H2020 Framework Program of the European Commission.
\end{acknowledgements}

\bibliographystyle{aa}
\bibliography{MassMapping}

\begin{thebibliography}{40}
\expandafter\ifx\csname natexlab\endcsname\relax\def\natexlab#1{#1}\fi

\bibitem[{Bacon {et~al.}(2006)Bacon, Goldberg, Rowe, \& Taylor}]{Bacon2006}
Bacon, D.~J., Goldberg, D.~M., Rowe, B. T.~P., \& Taylor, A.~N. 2006, Monthly
  Notices of the Royal Astronomical Society, 365, 414

\bibitem[{Bartelmann(1995)}]{Bartelmann1995}
Bartelmann, M. 1995, Astronomy and Astrophysics, 303, 643

\bibitem[{Bartelmann(2010)}]{Bartelmann2010}
Bartelmann, M. 2010, Classical and Quantum Gravity, 27, 233001

\bibitem[{Beckouche {et~al.}(2013)Beckouche, Starck, \& Fadili}]{Beckouche2013}
Beckouche, S., Starck, J.~L., \& Fadili, J. 2013, Astronomy and Astrophysics,
  556, A132

\bibitem[{Bradac {et~al.}(2004)Bradac, Lombardi, \& Schneider}]{Bradac2004}
Bradac, M., Lombardi, M., \& Schneider, P. 2004, Astronomy and Astrophysics,
  424, 13

\bibitem[{Bradac {et~al.}(2005)Bradac, Schneider, Lombardi, \&
  Erben}]{Bradac2005}
Bradac, M., Schneider, P., Lombardi, M., \& Erben, T. 2005, Astronomy and
  Astrophysics, 437, 39

\bibitem[{Cacciato {et~al.}(2006)Cacciato, Bartelmann, Meneghetti, \&
  Moscardini}]{Cacciato2006}
Cacciato, M., Bartelmann, M., Meneghetti, M., \& Moscardini, L. 2006, Astronomy
  and Astrophysics, 458, 349

\bibitem[{Cain {et~al.}(2015)Cain, Bradac, \& Levinson}]{Cain2015}
Cain, B., Bradac, M., \& Levinson, R. 2015, ArXiv e-prints, arXiv:1503.08218

\bibitem[{Cain {et~al.}(2011)Cain, Schechter, \& Bautz}]{Cain2011}
Cain, B., Schechter, P.~L., \& Bautz, M.~W. 2011, The Astrophysical Journal,
  736, 43

\bibitem[{Cand{\`{e}}s {et~al.}(2008)Cand{\`{e}}s, Wakin, \& Boyd}]{Candes2008}
Cand{\`{e}}s, E.~J., Wakin, M.~B., \& Boyd, S.~P. 2008, Journal of Fourier
  Analysis and Applications, 14, 877

\bibitem[{Chambolle \& Pock(2011)}]{Chambolle2011}
Chambolle, A. \& Pock, T. 2011, Journal of Mathematical Imaging and Vision, 40,
  120

\bibitem[{Clowe {et~al.}(2006)Clowe, Brada{\v{c}}, Gonzalez, Markevitch,
  Randall, Jones, \& Zaritsky}]{Clowe2006}
Clowe, D., Brada{\v{c}}, M., Gonzalez, A.~H., {et~al.} 2006, The Astrophysical
  Journal, 648, L109

\bibitem[{Condat(2013)}]{Condat2013}
Condat, L. 2013, Journal of Optimization Theory and Applications, 158, 460

\bibitem[{Deriaz {et~al.}(2012)Deriaz, Starck, \& Pires}]{Deriaz2012}
Deriaz, E., Starck, J.~L., \& Pires, S. 2012, Astronomy and Astrophysics, 540,
  A34

\bibitem[{Er {et~al.}(2010)Er, Li, \& Schneider}]{Er2010}
Er, X., Li, G., \& Schneider, P. 2010, ArXiv e-prints, arXiv:1008.3088

\bibitem[{Goldberg \& Bacon(2005)}]{Goldberg2005}
Goldberg, D.~M. \& Bacon, D.~J. 2005, The Astrophysical Journal, 619, 741

\bibitem[{Goldberg \& Leonard(2007)}]{Goldberg2007}
Goldberg, D.~M. \& Leonard, A. 2007, The Astrophysical Journal, 660, 1003

\bibitem[{Jullo {et~al.}(2014)Jullo, Pires, Jauzac, \& Kneib}]{Jullo2014}
Jullo, E., Pires, S., Jauzac, M., \& Kneib, J.~P. 2014, Monthly Notices of the
  Royal Astronomical Society, 437, 3969

\bibitem[{Kaiser \& Squires(1993)}]{Kaiser1993}
Kaiser, N. \& Squires, G. 1993, The Astrophysical Journal, 404, 441

\bibitem[{Keiner {et~al.}(2009)Keiner, Kunis, \& Potts}]{Keiner2009}
Keiner, J., Kunis, S., \& Potts, D. 2009, ACM Transactions on Mathematical
  Software, 36, 1

\bibitem[{Klypin {et~al.}(2011)Klypin, Trujillo-Gomez, \& Primack}]{Klypin2011}
Klypin, A.~A., Trujillo-Gomez, S., \& Primack, J. 2011, The Astrophysical
  Journal, 740, 102

\bibitem[{Leonard {et~al.}(2012)Leonard, Dup{\'{e}}, \& Starck}]{Leonard2012}
Leonard, A., Dup{\'{e}}, F.-X., \& Starck, J.-L. 2012, Astronomy {\&}
  Astrophysics, 539, A85

\bibitem[{Leonard {et~al.}(2007)Leonard, Goldberg, Haaga, \&
  Massey}]{Leonard2007}
Leonard, A., Goldberg, D.~M., Haaga, J.~L., \& Massey, R. 2007, The
  Astrophysical Journal, 666, 17

\bibitem[{Leonard \& King(2010)}]{Leonard2010}
Leonard, A. \& King, L.~J. 2010, Monthly Notices of the Royal Astronomical
  Society, 405, 1854

\bibitem[{Leonard {et~al.}(2009)Leonard, King, \& Wilkins}]{Leonard2009}
Leonard, A., King, L.~J., \& Wilkins, S.~M. 2009, Monthly Notices of the Royal
  Astronomical Society, 395, 1438

\bibitem[{Leonard {et~al.}(2014)Leonard, Lanusse, \& Starck}]{Leonard2014}
Leonard, A., Lanusse, F., \& Starck, J.-L. 2014, Monthly Notices of the Royal
  Astronomical Society, 440, 1281

\bibitem[{Massey {et~al.}(2015)Massey, Williams, Smit, Swinbank, Kitching,
  Harvey, Jauzac, Israel, Clowe, Edge, Hilton, Jullo, Leonard, Liesenborgs,
  Merten, Mohammed, Nagai, Richard, Robertson, Saha, Santana, Stott, \&
  Tittley}]{Massey2015}
Massey, R., Williams, L., Smit, R., {et~al.} 2015, Monthly Notices of the Royal
  Astronomical Society, 449, 3393

\bibitem[{Merten {et~al.}(2009)Merten, Cacciato, Meneghetti, Mignone, \&
  Bartelmann}]{Merten2009}
Merten, J., Cacciato, M., Meneghetti, M., Mignone, C., \& Bartelmann, M. 2009,
  Astronomy and Astrophysics, 500, 681

\bibitem[{Moreau(1962)}]{Moreau1962}
Moreau, J.-J. 1962, CRAS S{\'{e}}r. A Math., 255, 2897

\bibitem[{Okura {et~al.}(2007)Okura, Umetsu, \& Futamase}]{Okura2007}
Okura, Y., Umetsu, K., \& Futamase, T. 2007, The Astrophysical Journal, 660,
  995

\bibitem[{Paykari {et~al.}(2014)Paykari, Lanusse, Starck, Sureau, \&
  Bobin}]{Paykari2014}
Paykari, P., Lanusse, F., Starck, J.~L., Sureau, F., \& Bobin, J. 2014,
  Astronomy and Astrophysics, 566, A77

\bibitem[{Pires {et~al.}(2009)Pires, Starck, Amara, Teyssier, Refregier, \&
  Fadili}]{Pires2009}
Pires, S., Starck, J.~L., Amara, A., {et~al.} 2009, Monthly Notices of the
  Royal Astronomical Society, 395, 1265

\bibitem[{Rowe {et~al.}(2013)Rowe, Bacon, Massey, Heymans, Haussler, Taylor,
  Rhodes, \& Mellier}]{Rowe2013}
Rowe, B., Bacon, D., Massey, R., {et~al.} 2013, Monthly Notices of the Royal
  Astronomical Society, 435, 822

\bibitem[{Schneider \& Er(2008)}]{Schneider2008}
Schneider, P. \& Er, X. 2008, Astronomy and Astrophysics, 485, 363

\bibitem[{Seitz \& Schneider(1995)}]{Seitz1995}
Seitz, C. \& Schneider, P. 1995, Astronomy and Astrophysics, 297, 287

\bibitem[{Seitz \& Schneider(1996)}]{Seitz1996}
Seitz, S. \& Schneider, P. 1996, Astronomy and Astrophysics, 305, 383

\bibitem[{Starck {et~al.}(2015)Starck, Murtagh, \& Fadili}]{Starck2015}
Starck, J.-l., Murtagh, F., \& Fadili, J. 2015, {Sparse Image and Signal
  Processing: Wavelets and Related Geometric Multiscale Analysis} (Cambridge
  University Press)

\bibitem[{Starck {et~al.}(2006)Starck, Pires, \& Refregier}]{Starck2006}
Starck, J.~L., Pires, S., \& Refregier, A. 2006, Astronomy and Astrophysics,
  451, 1139

\bibitem[{Viola {et~al.}(2012)Viola, Melchior, \& Bartelmann}]{Viola2012}
Viola, M., Melchior, P., \& Bartelmann, M. 2012, Monthly Notices of the Royal
  Astronomical Society, 419, 2215

\bibitem[{Vu(2013)}]{Vu2013}
Vu, B.~C. 2013, Advances in Computational Mathematics, 38, 667

\end{thebibliography}

\newpage
\begin{appendix}

\section{Specialisation of the primal-dual algorithm}
\label{sec:appendix_prox}

This appendix introduces elements of proximal calculus and describes the specialisation  of the generic primal-dual algorithms introduced in \cite{Condat2013,Vu2013} to the specific weak lensing mass-mapping problems stated in this paper.

\subsection{Elements of proximal calculus}

Let $\mathcal{H}$ be a finite-dimensional Hilbert space (in the specific case considered in this paper, $\mathcal{H} = \mathbb{R}^n$). Let us consider a function $f: \mathcal{H} \rightarrow \mathbb{R} \cup \{ +\infty \}$ and define the \textit{domain of $f$}, noted $\dom f$, as the subset of $\mathcal{H}$ where $f$ does not reach $+ \infty$:
\begin{equation}
\dom f = \left\lbrace \bm{x} \in \mathcal{H} \quad | \quad f(\mathbf{x}) < + \infty \right\rbrace \;.
\end{equation}
A function $f$ will be said \textit{proper} if its domain $\dom f$ is nonempty. We also define the \textit{epigraph of $f$}, noted $\epi f$:
\begin{equation}
	\epi f = \left\lbrace (\bm{x},\lambda) \in \mathcal{H} \times \mathbb{R} \quad | \quad   f(\bm{x}) \leq \lambda \right\rbrace \;.
\end{equation}
The epigraph is useful to characterise several important properties, in particular the \textit{convexity} and \textit{lower semicontinuity} of $f$:
\begin{align}
\epi f \mbox{ is closed}  &\Leftrightarrow \mbox{$f$ is lower semicontinuous} \\
\epi f \mbox{ is convex} &\Leftrightarrow \mbox{$f$ is convex} 
\end{align}
We will note $\Gamma_0$ the class of proper lower semicontinuous convex functions of $\mathbb{R}^n$. Functions in $\Gamma_0$ are therefore characterised by a non empty closed convex epigraph.

The proximal operator, introduced by \cite{Moreau1962} is at the center of the different algorithms introduced in the following sections. This operator, which can be seen as an extension of the convex projection operator is defined by the  following:

\begin{definition}
Let $F \in \Gamma_0$. For every $\bm{x}$ the function $\bm{y} \mapsto \frac{1}{2} \parallel \bm{x} - \bm{y} \parallel^2 + F(\bm{y})$ achieves its infimum at a unique point defined as $\prox_{F}(\bm{x})$.  
\end{definition}

Therefore, given a proper lower semicontinuous convex function $F \in \Gamma_0$, the proximity operator of $F$ is uniquely defined by:
\begin{equation}
\prox_{F} (\bm{x}) = \argmin_{\bm{y}} \frac{1}{2} \parallel \bm{y} - \bm{x} \parallel_2^2 + F(\bm{y}) \;.
\label{eq:def_prox}
\end{equation}

In order to illustrate this operator in practice in a simple case, consider $F = i_{\mathcal{C}}$ the indicator function of a closed convex set $\mathcal{C}$. Then the proximity operator reduces to the orthogonal projector onto $\mathcal{C}$, noted $\proj_{\mathcal{C}}$:
\begin{equation}
	\prox_{i_{\mathcal{C}}}(\bm{x}) = \argmin_{\bm{y} \in \mathcal{C}} \frac{1}{2} \parallel \bm{y} - \bm{x} \parallel_2^2 = \proj_{\mathcal{C}}(\bm{x}) \;.
\end{equation}
Similarly to the simple case of the indicator function, explicit expressions for the proximity operator exist for a number of different simple functions. 

In the context of sparse optimisation, we will be particularly interested in the proximity operator of the $\ell_1$ norm. Thankfully, the proximity operator of $F(\bm{x}) = \lambda \parallel \bm{x} \parallel_1$ is explicit and corresponds to \textit{Soft Thresholding}:
\begin{equation}
	\prox_{\lambda \parallel \cdot \parallel_1} (\bm{x}) = \mathrm{ST}_{\lambda}(\bm{x}) \;,
\end{equation}
where the Soft Thresholding operator $\mathrm{ST}_{\lambda}$ is defined for $\bm{x} \in \mathbb{R}^N$ as:
\begin{equation}
	\forall i \in \llbracket1, N\rrbracket, \quad {\mathrm{ST}_{\lambda} (\bm{x})}_i = \left\lbrace \begin{matrix}
	x_i - \lambda &\mbox{ if $x_i \geq \lambda$} \\
	0   &\mbox{ if $|x_i| \leq \lambda$} \\
	x_i + \lambda & \mbox{ if $x_i \leq \lambda$}
\end{matrix}\right. \;.
\end{equation}

As previously  mentioned, the proximity operator is a key tool in convex optimisation as it can be used to derive fast convergent algorithms for convex but not  necessarily differentiable problems. In particular, in this paper we consider optimisation problems of  the following general form: 
\begin{equation}
	\argmin_{\bm{x}} F(\bm{x}) + G(W(\bm{x})) + H(\bm{x})
	\label{eq:general_minimisation}
\end{equation}
where $F$ is convex and differentiable, with a Lipschitzian gradient of  constant $\beta$, $(H,G) \in \Gamma_0^2$ and $W$ is a non-zero linear operator. In our case, $F$ will typically be the $\chi^2$ data fidelity term, $W$ the wavelet transform, and $G$ and $H$ the $\ell_1$ norm and the indicator function respectively, which are  convex but not differentiable.  Proximal algorithm generally rely on an iterative scheme using the gradient of the first term and the proximity operators of the additional terms. The particular difficulty of this problem stems from the composition of the two operators $G$ and $W$ which does not possess in the  general case an explicit proximity operator even if $\prox_{G}$ has a closed-form expression.

Primal-dual algorithms, such as the one initially proposed in \cite{Chambolle2011}, avoid this difficulty by recasting part of the problem in a form which allows the separation between $G$ and $W$ such that a closed-form expression of $\prox_G$ can be used if it exists. The main  limitation of this algorithm however is that it does not take advantage of the differentiability of the first term $F$. More recently, a variation of this approach was proposed in \cite{Condat2013,Vu2013}, still relying on a primal-dual formulation for  separating $G$ and $W$ but taking full advantage of the differentiability of $F$, thus leading to a much more efficient algorithm in practice. The main iteration of this minimisation algorithm for solving \autoref{eq:general_minimisation} takes the following form:
\begin{equation}
\left\lbrace\begin{matrix}
 \bm{x}^{(k+1)} &=& \prox_{\tau H} \left(  \bm{x}^{(k)}  + \tau (\nabla F( \bm{x}^{(k)}) - W^* \bm{u}^{(k)} ) \right)\\
 \bm{u}^{(k+1)} &=& \prox_{G^*} \left(  \bm{u}^{(k)}  + \sigma W (2 \bm{x}^{(k+1)}  - \bm{x}^{(k)}) \right)
\end{matrix}\right.
\label{eq:primal_dual_main_iteration}
\end{equation}
where $\tau$ and $\sigma$ must verify $1 - \tau \sigma \parallel W \parallel^2 > \tau \beta/2$ to ensure convergence and where $G^*$ is the convex conjugate of the original function $G$, defined according to $G^*(\bm{x}) = \max_{\bm{y}} \{ <\bm{y}, \bm{x}> - G(\bm{y})\} $. Thankfully, the proximal operator of the convex conjugate is readily expressible in terms of the proximal operator of the original function, with $\prox_{G^*}(\bm{x}) = \bm{x} - \prox_G(\bm{x})$, so that the proximity operator of the conjugate  can be evaluated  at no extra-cost.

\subsection{Algorithm specialisation in the  simplified linear setting}
\label{sec:appendix_linear_problem}

In the simplified linear setting, deriving the specialisation of minimisation algorithm introduced in \autoref{eq:primal_dual_main_iteration} is straightforward. We aim to solve \autoref{eq:conv_sparse_rec_lin}, which can be seen as a specialisation of \autoref{eq:general_minimisation} for  the following choice of operators:
\begin{align}
	F(\bm{\kappa}) &= \frac{1}{2}\parallel \Sigma^{-\frac{1}{2}}_{\gamma} \left[ \bm{\gamma} - \mathbf{T} \mathbf{P} \mathbf{F}^* \bm{\kappa} \right] \parallel_2^2  &;&\quad G(\bm{\alpha}) &=&  \lambda \parallel \bm{w} \circ \bm{\alpha} \parallel_1 \nonumber\\  W(\bm{\kappa}) &= \mathbf{\Phi}^* \bm{\kappa}  &;& \quad H(\bm{\kappa}) &=& i_{\Im(\cdot) = 0}(\bm{\kappa})\nonumber
\end{align}
Applying the \cite{Condat2013,Vu2013} algorithm introduced in the previous section is  therefore only  a matter of computing  the gradient of $F$ and  the proximity operators of $G^*$ and $H$. The gradient of the quadratic term $F$ is simply:
\begin{equation}
	\nabla F( \bm{\kappa})  = \mathbf{F} \mathbf{P}^* \mathbf{T}^* \Sigma_\gamma^{-1} \left(\bm{\gamma} -  \mathbf{T} \mathbf{P} \mathbf{F}^{*} \bm{\kappa} \right)
\end{equation}
 while the proximity operators of $G$ and $H$ have a closed-form expression already introduced in the previous section:
 \begin{align}
 	\prox_{G^*}(\bm{\alpha}) &= \bm{\alpha} - \mathrm{ST}_{\lambda \circ \bm{w}}( \bm{\alpha}) \\
 	\prox_{H}(\bm{\kappa}) &= \Re(\bm{\kappa})
 \end{align}
These expressions for the gradient and proximity operators readily yield \autoref{alg:2Dmassmap_lin} when applied to the iteration introduced in \autoref{eq:primal_dual_main_iteration}.

\subsection{Algorithm specialisation for the complete problem}
\label{sec:appendix_full_problem}

Contrary to the previous case, solving the full mass-mapping problem stated in \autoref{eq:full_conv_optim} presents a number of difficulties. The specialisation  of \autoref{eq:general_minimisation} differs from the previous  case in the definition of the differentiable term $F$ and the convex function $H$:
\begin{align}
	F\left(\left[ \begin{matrix}
		 \bm{\kappa}  \\
		 \bm{\tilde{\mathcal{F}}}
\end{matrix}\right] \right) &= \frac{1}{2} \parallel \mathcal{C}_{\kappa g}^{-1} \left[ (1 - \bm{Z} \mathbf{T} \mathbf{F}^* \bm{\kappa}) \bm{g} - \bm{Z}\mathbf{T} \mathbf{P} \mathbf{F}^* \bm{\kappa} \right] \parallel_2^2 \nonumber \\
 &  + \frac{1}{2} \parallel \mathcal{C}_{\kappa F}^{-1} \left[ (1 - \bm{Z} \mathbf{T} \mathbf{F}^* \bm{\kappa}) \bm{F} - \bm{Z} \mathbf{T} \mathbf{F}^* \tilde{\bm{\mathcal{F}}} \right] \parallel_2^2 \nonumber \\
 H\left(\left[ \begin{matrix}
		 \bm{\kappa}  \\
		 \bm{\tilde{\mathcal{F}}}
\end{matrix}\right] \right) &= i_{\mathrm{Im}(\mathbf{R})}\left( \left[ \begin{matrix}
\bm{\kappa} \\
\tilde{\bm{\mathcal{F}}}
\end{matrix}	\right]  \right) \nonumber
\end{align}

Remember that this expression for the quadratic fidelity term $F$ is a linearisation of the original problem which includes a non-linear factor due to the  reduced shear. However, despite  this first simplification, the gradient of this expression proves to be problematic. Since the shear and flexion terms are equivalent and separable, let us concentrate on the gradient of the shear term of $F$, which takes the form:
\begin{multline}
\mathbf{F} \mathbf{P}^* \mathbf{T}^*  \mathbf{Z} \mathcal{C}_{\kappa g}^{-2} \left[ (1 - \mathbf{Z}  \mathbf{T} \mathbf{F}^* \bm{\kappa}) \bm{g} -  \mathbf{Z} \mathbf{T} \mathbf{P} \mathbf{F}^* \bm{\kappa} \right] \\  + \mathbf{F} \mathbf{T}^* \mathbf{Z}  \bm{g}^* \mathcal{C}_{\kappa g}^{-2} \left[ (1 - \mathbf{Z}  \mathbf{T} \mathbf{F}^* \bm{\kappa}) \bm{g} -  \mathbf{Z} \mathbf{T} \mathbf{P} \mathbf{F}^* \bm{\kappa} \right]
 \label{eq:gradient}
\end{multline} 
The first term of this expression simply corresponds to the same gradient as in the linear case but corrected for the reduced shear. In particular, the noise affecting the measured shear still propagates linearly. On the contrary, the second term is new compared to the linear case and is problematic in the sense that it becomes a quadratic function of the reduced shear $\bm{g}$. If one assumes the reduced shear noise to be Gaussian, the noise contribution of this second term becomes $\chi^2$ distributed. As explained in \autoref{subsec:sparse_constraint}, our regularisation scheme is defined in terms of the standard deviation of the noise propagated to the wavelet coefficients. While this scheme proves very effective for Gaussian noise, it is not appropriate for $\chi^2$ distributed noise.

Therefore, in the approach presented here, we choose to use a suboptimal gradient for the quadratic data fidelity term by keeping only the first term of \autoref{eq:gradient}, which is the price to pay in order to keep the very effective regularisation scheme introduced in the linear case. Thus, we use the following expression instead of the exact gradient of $F$:
\begin{align}
\nabla F &= \mathbf{F}\mathbf{P}^* \mathbf{T}^* \mathbf{Z} \mathcal{C}_{\kappa g}^{-2} \left( (1 - \mathbf{Z} \mathbf{T} \mathbf{F}^* \bm{\kappa}) \bm{g} -  \mathbf{Z}\mathbf{T} \mathbf{P} \mathbf{F}^* \bm{\kappa} \right) \nonumber \\ 
 &+ \mathbf{F} \mathbf{T}^*  \mathbf{Z}\mathcal{C}_{\kappa F}^{-2} \left( (1 -  \mathbf{Z}\mathbf{T} \mathbf{F}^* \bm{\kappa}) \bm{F} -  \mathbf{Z}\mathbf{T} \mathbf{F}^* \tilde{\bm{\mathcal{F}}}\right) \nonumber
\end{align}
 Note however that despite this simplification, the resulting procedure still corresponds to the reduced shear correction scheme suggested in \cite{Seitz1995}.

\bigskip

The second difference with the shear only reconstruction comes from the more complex constraint $H$ based on the indicator function of the image of the linear operator $\mathbf{R}$ introduced in \autoref{eq:image_of_R}. Thankfully, despite the additional complexity, the  proximity operator of this term can still be explicitly from its definition:
\begin{align}
	\prox_{i_{\mathrm{Im}(\mathbf{R})}}\left(\left[ \begin{matrix}
	\bm{\kappa} \\
	\tilde{\bm{\mathcal{F}}}
\end{matrix}	\right] \right) &= \argmin_{(\bm{x}_{\kappa}, \bm{x}_{\mathcal{F}})} \frac{1}{2} \parallel \left[ \begin{matrix}
	\bm{\kappa} \nonumber \\
	\tilde{\bm{\mathcal{F}}}
\end{matrix}	\right]  - \left[ \begin{matrix}
	\bm{x}_{\kappa} \\
	\bm{x}_{\mathcal{F}}
\end{matrix}	\right] \parallel_2^2 + i_{\mathrm{Im}(\mathbf{R})} \left( \left[ \begin{matrix}
	\bm{x}_{\kappa} \\
	\bm{x}_{\mathcal{F}}
\end{matrix}	\right] \right)  \\
		&= \mathbf{R} \argmin_{\bm{x}_{\kappa} \in \mathbb{R}^{N \times N}}  \parallel \left[ \begin{matrix}
	\bm{\kappa} \\
	\tilde{\bm{\mathcal{F}}}
\end{matrix}	\right]  - \left[ \begin{matrix}
	\bm{x}_{\kappa} \\
	\mathbf{F} \mathbf{Q} \mathbf{F}^* \bm{x}_{\kappa}
\end{matrix}	\right] \parallel_2^2\nonumber \\
	& = \mathbf{R} \Re \left( \mathbf{F} \frac{1}{k^2 + \frac{\sigma_\mathcal{F}^2}{\sigma_\epsilon^2}} \left( k^2 \mathbf{Q}^{-1} \mathbf{F}^* \tilde{\bm{\mathcal{F}}} + \frac{\sigma_\mathcal{F}^2}{\sigma_\epsilon^2} \mathbf{F}^* \bm{\kappa} \right) \right)\nonumber  \;.
\end{align}
In this expression, $\sigma_\epsilon^2$ is the variance of the intrinsic ellipticity and $\sigma_\mathcal{F}^2$ is the variance of the intrinsic flexion. Let us detail what is computed by this proximity operator. The first step is to rewrite the proximity operator using its definition, introduced in \autoref{eq:def_prox}. Then, using the  definition of $\mathrm{Im}(\mathbf{R})$, the problem can be recast in terms of a single real unknown $\bm{x}_\kappa$, from which convergence and flexion can be computed by applying the operator $\mathbf{R}$. Finally, as $\bm{\kappa}$ is fitted to the shear and $\tilde{\bm{\mathcal{F}}}$ is fitted to the flexion, solving the minimisation problem for $\bm{x}_\kappa$ amounts to finding the minimum variance filter combining shear and flexion, already introduced in \autoref{eq:flexion_minimum_variance}. To summarise, this operator computes the optimal combination of the two input variables, making explicit use of the flexion Fourier operator and its inverse $\mathbf{Q}^{-1} = \mathbf{Q}^*/k^2$, thus eliminating the need to solve this additional problem as part of the main minimisation problem.

Given these expressions for  $\nabla F$ and $\prox_{i_{\mathrm{Im}(\mathbf{R})}}$, the specialisation of the primal-dual algorithm yields \autoref{alg:2Dmassmap_full}.

\end{appendix}
\end{document}